\documentclass[prb,amsmath,amssymb,superscriptaddress,twocolumn]{revtex4}
\usepackage{graphicx} 
\usepackage{color}

\newcommand{\br}{{\bf r}}
\newcommand{\bx}{{\bf x}}
\newcommand{\by}{{\bf y}}
\newcommand{\bk}{{\bf k}}

\newcommand{\bp}{{\bf p}}

\newcommand{\bz}{{\bf z}}
\newcommand{\bq}{{\bf q}}
\newcommand{\bA}{{\bf A}}

\newcommand{\bB}{{\bf B}}
\newcommand{\bK}{{\bf K}}

\newcommand{\bR}{{\bf R}}

\newcommand{\piv}{\mbox{\boldmath$\pi$}}

\DeclareMathAlphabet{\mathpzc}{OT1}{pzc}{m}{it} 

\DeclareMathOperator{\realpart}{Re}
\begin{document}
\title{Fermions on bilayer graphene: symmetry breaking for $B=0$ and $\nu=0$.}
\author{Robert E. Throckmorton}
\author{Oskar Vafek}
\affiliation{National High Magnetic Field Laboratory and Department
of Physics, Florida State University, Tallahassee, Florida 32306,
USA}
\date{\today}
\begin{abstract}
We extend previous analyses of fermions on a honeycomb bilayer
lattice via weak-coupling renormalization group (RG) methods with
extremely short-range and extremely long-range interactions to the
case of finite-range interactions. In particular, we consider
different types of interactions including screened Coulomb
interactions, much like those produced by a point charge placed
either above a single infinite conducting plate or exactly halfway
between two parallel infinite conducting plates. Our considerations
are motivated by the fact that, in some recent experiments on
bilayer graphene\cite{YacobyPRL2010, Mayorov2011} there is a single
gate while in others\cite{YacobyScience2010, VelascoNatNano2012}, there
are two gates, which can function as the conducting planes and which, we
argue, can lead to distinct broken symmetry phases. We map out the
phases that the system enters as a function of the range of the
interaction. We discover that the system enters an antiferromagnetic
phase for short ranges of the interaction and a nematic phase at long
ranges, in agreement with Refs.\ \onlinecite{VafekYangPRB2010} and
\onlinecite{VafekPRB2010}.  While the antiferromagnetic phase results
in a gap in the spectrum, the nematic phase is gapless, splitting the
quadratic degeneracy points into two Dirac cones each\cite{VafekPRB2010, LemonikPRB2010, LemonikArXiv}.
We also consider the effects of an applied magnetic field on the system
in the antiferromagnetic phase via variational mean field theory.  At
low fields, we find that the antiferromagnetic order parameter, $\Delta(B)-\Delta(0)\sim B^2$.
At higher fields, when $\omega_c\gtrsim 2\Delta_0$, we find that
$\Delta(B)\approx\omega_c/[\ln(\omega_c/\Delta(0))+C]$,
where $C\approx 0.67$ and $\omega_c=eB/m^*c$. We also determine the energy
gap for creating electron-hole excitations in the system, and, at high fields, we
find it to be $a\omega_c+2\Delta(B)$, where $a$ is a non-universal, interaction-
dependent, constant.
\end{abstract}
\maketitle
\section{Introduction}
The problem of interacting fermions on an A-B stacked honeycomb
bilayer lattice has been of great interest, both theoretically and
experimentally, due in no small part to its band structure.
Typically, the band structure possesses two inequivalent points,
labeled $\bK$ and $\bK'=-\bK$, in the Brillouin zone at which the
two low-energy bands make contact at four Dirac-like
cones\cite{CastroNetoRMP2009}.  However, in the case where we
neglect all but nearest-neighbor hopping terms, these four Dirac
cones merge into a single quadratic degeneracy
point\cite{CastroNetoRMP2009}.  The density of states at the Fermi
level at half-filling therefore becomes finite in this special case.
This leads to logarithmic divergences in the uniform and static
susceptibilities to various symmetry-breaking orders at zero
temperature\cite{VafekPRB2010, VafekYangPRB2010}, and thus we expect
such phases to appear for arbitrarily weak interactions over some
finite range of temperatures.

One key motivation for the study of the honeycomb bilayer in
particular is the fact that bilayer graphene, a material of great
interest, possesses this lattice structure.  In particular,
measurements of the conductivity of suspended bilayer graphene in
various external electric and magnetic fields as well as different
carrier densities\cite{YacobyScience2010}, followed by
compressibility measurements\cite{YacobyPRL2010}, provide evidence
suggesting that a symmetry-breaking state appears; the two possible
states argued for in these works are a quantum anomalous Hall
phase, in which the system develops a non-zero quantized Hall
conductivity in the absence of a magnetic field, and a nematic
phase, in which the quadratic degeneracy points each split into two
massless Dirac-like cones.  A more recent
experiment\cite{Mayorov2011} finds evidence for the presence of a
nematic phase\cite{VafekYangPRB2010,LemonikPRB2010,LemonikArXiv} by measuring the
width of a peak in the resistivity as a function of the carrier
density at different temperatures and by measuring cyclotron gaps as
a function of the applied magnetic field for different filling
factors. Another, even more recent, experiment\cite{VelascoNatNano2012} uses
measurements of two-terminal conductance to argue for a state in
which the system develops an energy gap in its spectrum, in apparent
disagreement with the previous experiment, since the nematic state
is gapless.  Two other experiments\cite{FreitagPRL2012,VeliguraPRB2012} also conclude
that a gap due to symmetry breaking is present, but are inconclusive about the exact
nature of the state.  Yet another experiment\cite{BaoArXiv}
performs measurements on multiple samples, finding a bimodal distribution
of both conducting and insulating samples.

In addition to experimental investigations, there have also been a
number of theoretical studies regarding the nature of the
symmetry-breaking states of the A-B stacked honeycomb bilayer.  One
such investigation\cite{NilssonPRB2006} uses variational methods to
argue for a ferromagnetic phase for long-range interactions and
a calculation of the susceptibility towards an antiferromagnetic phase
within an RPA approximation, followed by a mean-field calculation of the
associated order parameter, to argue for said phase for short-range interactions.
Two other works employ a mean-field approach to argue for a ``(layer) pseudospin
magnet'' state\cite{MinPRB2008} and a ferromagnetic state\cite{CastroPRL2008}, while later
investigations argue for a layer-polarized state\cite{NandkishorePRL2010}
within the mean-field approximation and a quantum anomalous Hall
state\cite{NandkishorePRB2010} with mean field plus Gaussian fluctuations.
Another work by the same group\cite{NandkishoreArXiv} considers the phase
of the system within the mean-field approximation as a function of external
electric and magnetic fields and finds a quantum anomalous Hall
state, a quantum Hall ferromagnetic insulator, and a layer-polarized
state.  Another work\cite{JungPRB2011} uses Hartree-Fock methods to argue
for a layer-polarized state in the absence of a magnetic field, and argues
for the existence of an anomalous quantum Hall state in the presence of a
magnetic field.  A very recent investigation by the same group arrives
at a ``(layer) pseudospin antiferromagnetic'' state for the honeycomb
bilayer\cite{MacDonaldPhysScr2012}.  Another very recent investigation finds,
using Hartree-Fock methods, a coexistence of a quantum spin Hall state and
a layer-polarized state that may be turned into a pure layer-polarized state
with the application of a sufficiently strong electric field\cite{GorbarPRB2012}.
The work of Ref.\ \onlinecite{ZhangPRL2012} treats the problem of the combined
effect of an in-plane magnetic field, i.e., with a Zeeman term only, and an applied
perpendicular electric field, within a self-consistent mean field approximation.

The mean-field approach, {\it as a means of predicting the low-temperature
phase of the system}, however, has one major disadvantage.  It does not treat
the leading logarithmic divergences that appear in perturbation theory correctly,
and therefore may not lead to results that are correct, even qualitatively\cite{DzyaloshinskiiLarkin1972}.
On the other hand, the renormalization group (RG) approach does.  An example of
this point is provided by the treatment of interacting fermions on a one-dimensional
chain\cite{DzyaloshinskiiLarkin1972}.  In the spinless case (see,
for example, Ref.\ \onlinecite{ShankarRMP1994}), mean field theory predicts
that there will be a charge density wave state for arbitrarily weak
interactions at half filling, while RG predicts a Luttinger liquid for weak
interactions.  We therefore believe that RG is more accurate as a
means of predicting the state that our system enters than a mean
field approach.  There are now several papers that employ the RG
approach in studying the honeycomb bilayer in the weak-coupling
limit\cite{VafekPRB2010, VafekYangPRB2010, LemonikPRB2010, LemonikArXiv,
MacDonaldPRB2010}.

A paper by one of us and Yang\cite{VafekYangPRB2010} uses this
method to argue for the existence of a nematic state for extremely
long-range interactions in the case of spin-$\tfrac{1}{2}$ fermions,
and a similar paper by Lemonik \textit{et. al.}\cite{LemonikPRB2010,LemonikArXiv}
which followed comes to the same conclusion.  Another paper by one
of us\cite{VafekPRB2010} investigates the extremely short-range
limit, namely the repulsive Hubbard model, using the same approach,
and finds that a system of spin-$\tfrac{1}{2}$ fermions will enter an antiferromagnetic state;
it is also argued within that this state should persist even
in the strong-coupling limit.  This has been very recently confirmed\cite{LangArXiv}
using a combination of quantum Monte Carlo and functional RG methods.

In the present work, we will extend the analyses conducted in
Refs.\ \onlinecite{VafekYangPRB2010} and \onlinecite{VafekPRB2010} to the case of
finite-ranged density-density interactions in the case of spin-$\tfrac{1}{2}$
fermions, and to finite temperature, following the methods described in detail
in Ref.\ \onlinecite{CvetkovicArXiv}.  We are interested in determining how the
system transitions from the antiferromagnetic state to the nematic state as we
increase the range of the interaction.  In particular, we consider two
types of interactions, namely 1) screened Coulomb interactions, much like
those produced by a point charge placed exactly halfway between two parallel
infinite conducting plates, 2) same as 1) but for a single infinite conducting
plate. Each form considered includes a parameter $\xi$ that is proportional to
the range of the interaction.

Our procedure for the calculation is as follows. We start with a
tight-binding lattice model for the honeycomb bilayer with a
density-density interaction between electrons.  From this, we may
derive a low-energy effective field theory, which takes the same
form as that given in the previous work\cite{VafekPRB2010}; said
field theory contains 9 different contact quartic couplings that are
allowed by the symmetries of the honeycomb bilayer lattice and by
Fierz identities. We find the values of the coupling constants in
terms of the interaction in our original tight-binding lattice
microscopic model.  We then use these values as the initial
conditions for the RG flow equations derived to one-loop order in
Ref.\ \onlinecite{CvetkovicArXiv}, which we integrate numerically.
We then perform the analysis outlined therein to determine the phase
that the system enters.

For very short-ranged interactions, we find an antiferromagnetic phase, in
agreement with the previous work\cite{VafekPRB2010}.  This is determined by
monitoring the susceptibilities toward various phases as the temperature is
lowered; in this case, the only susceptibility that diverges at the critical
temperature is towards the antiferromagnetic phase.  If we increase the
range, we will enter a region where the susceptibilities toward both the antiferromagnetic
and nematic phases both diverge, although not necessarily with the same exponent\cite{CvetkovicArXiv}.
This transition occurs for values of $\xi$ anywhere from about $0.4$ to $2$
lattice spacings, depending on the form and overall strength of the interaction.
Finally, upon increasing $\xi$ further, we enter a pure nematic phase, again in
agreement with the previous work\cite{VafekYangPRB2010,VafekPRB2010}.  This
happens for values of $\xi$ anywhere from $4$ to about $10$ lattice spacings,
again depending on the form and overall strength of the interaction.

For the dipole-like potential, much like the one that is produced by
a point charge a distance $\xi$ above a single infinite conducting
plate,
\begin{equation}
V(\br)=U_0\left [\frac{1}{r/\xi}-\frac{1}{\sqrt{(r/\xi)^2+1}}\right ], \label{Eq:ScrCoul_OP_Intro}
\end{equation}
where $U_0$ is a sufficiently small constant, so that our system is
at weak coupling. For the potential produced by a point charge
placed exactly half-way between two infinite conducting plates separated
by a distance $2\xi$,
\begin{equation}
V(\br)=4U_0\sum_{k=0}^{\infty}K_0\left [(2k+1)\pi\frac{r}{\xi}\right
]\approx U_0\frac{2\sqrt{2}e^{-\pi r/\xi}}{\sqrt{r/\xi}}, \label{Eq:ScrCoul_TP_Intro}
\end{equation}
where $K_n(x)$ is a modified Bessel function of the second kind and the last equation
holds asymptotically at large values of $r/\xi$.  Note that these expressions,
as they are, diverge at $\br=0$---that is, they diverge for two electrons on the
same lattice site.  Because of this, and due to the fact that the electrons in the
tight-binding model cannot, strictly speaking, be considered point particles, we render
these on-site interactions finite by setting said interactions equal to a
constant $\lambda$ times the nearest-neighbor interaction; our main results
were found for $\lambda=1.2$; this choice makes our interactions monotonically decreasing
with distance.

In each case, we find a qualitatively identical phase diagram, except
that the values of $\xi$ at which the transitions described earlier
occur are smaller for the one-plate case than for the two-plate case.

Another way to think of our results is as follows.  When we make our
interactions longer-ranged in real space, we are, at the same time,
making them shorter-ranged in momentum space.  Let us start with a
density-density interaction of the general form,
$\sum_{\br\br'}V(\br-\br')n(\br)n(\br')$, where
$n(\br)=\sum_{\sigma}c_{\sigma}^{\dag}(\br)c_{\sigma}(\br)$.  It
turns out that the initial couplings in the effective low-energy
field theory depend on the Fourier components $V(\bq)$ of this
interaction at $\bq=0$ and $\bq=2\bK$; we will label these
components as $V_0$ and $V_{2K}$, respectively.  In the case of a
long-range interaction $V_0\gg V_{2K}$, and said interaction will
therefore mostly cause forward scattering of the electrons while, in
the case of a short-range interaction, $V_{2K}$ will become
comparable to $V_0$, and thus back scattering begins to become
comparable to the forward scattering.  We would therefore say that
we obtain a nematic phase when forward scattering dominates, while
we obtain an antiferromagnetic phase when back scattering becomes
comparable to the forward scattering.

Interestingly, if we make the on-site (repulsive) interaction \textit{weaker}
than the nearest-neighbor (repulsive) interaction, we may also obtain a quantum
spin Hall phase for intermediate ranges of the interaction; whether
or not it appears depends on the sign of one of our coupling
constants, $g_{E_K}$, which we define in the main text, for a given
value of $\xi$.  This may be seen from Fig.\ 5 in Ref.\ \onlinecite{CvetkovicArXiv}.
We were, however, unable to obtain a quantum anomalous Hall state
in this case with any density-density interaction.

Motivated by experimental data\cite{YacobyPRL2010,VelascoNatNano2012} and by our above
conclusions, we then turn our attention to investigating the effects
of an applied magnetic field on the antiferromagnetic phase.  A
similar calculation has already been performed\cite{KharitonovArXiv},
in which a self-consistent mean field approach is employed, and a second
order parameter corresponding to a ``staggered spin current'' order\cite{CvetkovicArXiv}
is explicitly introduced.  Our variational mean-field analysis only explicitly includes
the antiferromagnetic order parameter.  We do this because, while at $B=0$,
the two order parameters are distinct in that they have opposite parity under
mirror reflections, a finite external magnetic field, which is an axial vector,
breaks the mirror symmetry.  Therefore, at $B\neq 0$, the two order
parameters automatically mix; they belong to the same ($A_{1u}$) 
representation of $S_6$, the point group for $B\neq 0$.  
Indeed, our variational wave function, which
includes only the AF order parameter explicitly, leads to a finite expectation
value of the ``staggered spin current'' order parameter when $B\neq 0$.  In
Ref.\ \onlinecite{KharitonovArXiv}, the development of a finite expectation
value of the ``staggered spin current'' operator was attributed to ``the
emergence of the $n=0,1$ Landau levels (LLs) and the peculiar property of
their wave-functions to reside on only one sublattice in each valley''.  Here,
we extend the argument and show that it must be present on more general
grounds, and is not tied to the properties of the Landau levels. Note that,
in this case, we are using mean-field methods to determine the phenomenology
of the broken-symmetry state in a case where we already know which symmetries are
broken, based on our RG calculations, rather than as a means of {\it predicting}
the broken-symmetry state.  In fact, since the AF state is gapped, we expect that an
expansion around the mean-field state should have a finite radius of convergence. 
Another reason why we employ mean-field methods rather than RG in this calculation is because,
in the presence of a perpendicular magnetic field, the non-interacting energy
spectrum is discrete and momentum $\bk$ is not a good quantum number, 
thus making an analysis of the type used previously more difficult.

In order to obtain a fit to the experimentally-measured gap\cite{YacobyPRL2010,VelascoNatNano2012},
we calculate the energy required to create a (charge neutral) particle-hole
excitation, from which we can find the energy gap in the system.  We find that it is
\begin{equation}
E_{\text{ex}}=\min[E_{n_1}+E_{n_2}+\frac{m^*}{4\pi}g^*\omega_c(\tau_1 s_1-\tau_2 s_2)], \label{Eq:Eex_Intro}
\end{equation}
where $\omega_c=eB/m^*c$ is the cyclotron frequency for the effective mass $m^*$, and
\begin{equation}
E_n=\sqrt{n(n-1)\omega_c^2+\Delta(B)^2}.
\end{equation}
$\tau$ and $s$ are valley and spin indices; the former is $\pm1$ for the $\pm\bK$ valley, and
the latter is $\pm1$ for spin up (down).  Note that, if $n_j=0$ or $1$, then $\tau_js_j$ is
locked to $+1$ for $j=1$ (particle) and to $-1$ for $j=2$ (hole).  This is simply
the sum of the energies of the two states in the single-particle ``auxiliary spectrum'',
plus a non-universal term linear in the applied magnetic field, which depends on the coupling
constant $g^*\sim V_0-V_{2K}$ discussed further in the text.  The gap in our system is simply the
minimum energy required to create such a particle-hole excitation.  As discussed in the
next paragraph, while the field dependence of the AF order parameter $\Delta(B)$ is
universally determined by $\Delta(0)$ through Eq.\ \eqref{Eq:AF_SC_Intro}, the energy gap is not.
The $B$-linear term has a coefficient that depends on the interaction in the combination, $V_0-V_{2K}$,
and thus the slope of the energy gap at high fields is independent of the zero-field
value of the order parameter $\Delta_0=\Delta(0)$.  These results are in agreement with those
found earlier in Ref.\ \onlinecite{KharitonovArXiv}.

In our method of determining $\Delta(B)$, which was subsequently reproduced in 
Ref.\ \onlinecite{KharitonovArXiv}, we eliminate the coupling constants in the problem
by rewriting the self-consistent equation in terms of $\Delta_0$.  This allows us to send the
energy cutoff in our problem to infinity. By doing so, we can obtain
the scaling equation for the order parameter at a finite magnetic field {\it
whose dependence on coupling constants enters entirely through} $\Delta_0$:
\begin{equation}
F\left (\frac{\Delta(B)}{\omega_c}\right )=\ln\left (\frac{\omega_c}{\Delta_0}\right ), \label{Eq:AF_SC_Intro}
\end{equation}
where
\begin{equation}
F(\alpha)=I(\alpha)+\frac{1}{|\alpha|}-\ln\left
(1+\sqrt{\alpha^2+\tfrac{3}{4}}\right )
\end{equation}
and $I(\alpha)$ is stated in Eq.\ \eqref{Eq:I_Integral},
but may be very accurately approximated for {\it any} value of $\alpha$ as
\begin{equation}
I(\alpha)\approx -\frac{2}{[(-\tfrac{1}{2}I(0))^{-2/3}+4\cdot
6^{2/3}\alpha^2]^{3/2}},
\end{equation}
and $I(0)\approx -0.0503$ is the exact value of $I(\alpha)$ at
$\alpha=0$.

While the above equation must, in general, be solved for $\Delta(B)$
numerically, it is possible to obtain approximate analytic solutions
in certain limits. For low fields, the order parameter has a quadratic
dependence on the magnetic field,
\begin{equation}
\Delta(B)=\Delta_0+\frac{\omega_c^2}{8\Delta_0},
\end{equation}
while, at high magnetic fields, the dependence on the field is
\begin{equation}
\Delta(B)=\frac{\omega_c}{\ln(\omega_c/\Delta_0)+C}, \label{Eq:Delta_HF_Intro}
\end{equation}
where $C\approx 0.67$. This may be experimentally hard to
distinguish from linear dependence.

We find that $E_{\text{ex}}$, given by Eq.\ \eqref{Eq:Eex_Intro}, has a slight
non-monotonic behavior for small fields.  For example, if we choose $\frac{m^*}{4\pi}g^*=0.44$
and $\Delta_0=0.95\text{ meV}$, which we use to fit the experiment in Ref.\ \onlinecite{VelascoNatNano2012},
we find a minimum of about $1.91\Delta_0$, occuring at around $B=0.047\text{ T}$.  At larger
fields, we observe a ``kink'', marking a transition to an approximately linear
behavior, at a field of about $0.45\text{ T}$.  This may be seen in Fig.\ \ref{Fig:EXSpecPlot}.

We also considered the high-field data for the $\nu=0$ gap given in Ref.\ \onlinecite{YacobyPRL2010}.
In this case it is unclear from the low-field data what is the value of the energy
gap at zero applied field, if any.  Nevertheless, because of the weak logarithmic
dependence on $\Delta_0$ shown in Eq.\ \eqref{Eq:Delta_HF_Intro}, it is reasonable to
fit this data using the expressions presented above and the same $\Delta_0=0.95\text{ meV}$
as for the fit to the data of Ref.\ \onlinecite{VelascoNatNano2012}.  In this case,
we still obtain a non-monotonic dependence on the field, but with a very
shallow minimum at about $B=0.017\text{ T}$.  In this case, we would need to assume
$V_0<V_{2K}$, which would require a non-monotonic density-density interaction.

The rest of the paper is organized as follows.  In Section II,
we state the starting microscopic lattice Hamiltonian.  In Section III,
we derive the corresponding low-energy effective theory and
the relations between the coupling constants in the low-energy theory
and the interaction in the microscopic lattice Hamiltonian.  Section IV
is dedicated the determination of the phase of the system as a function
of the range of the interaction using the results of Section III as the
initial conditions for the RG analysis detailed in Ref.\ \onlinecite{CvetkovicArXiv}.
In Section V, we present our variational mean field analysis of the
antiferromagnetic state in the presence of an applied magnetic
field and calculation of the energy gap within this approximation.
We present our conclusions in Section VI, and provide mathematical details
in the appendix.

\section{Microscopic lattice Hamiltonian}
\begin{figure}[tb]
\centering
\includegraphics[width=\columnwidth,clip]{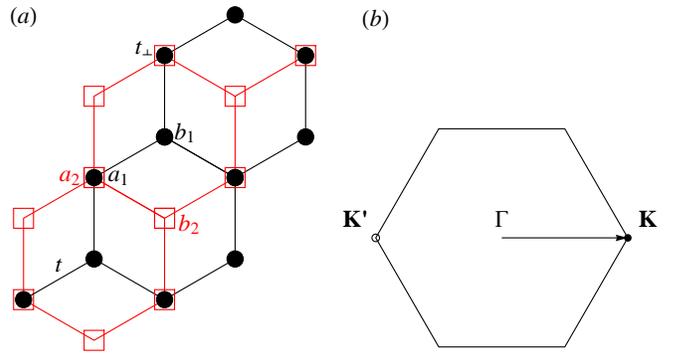}
\caption{\label{HCBilayer_Diagram}$(a)$ Illustration of the honeycomb
bilayer lattice with only nearest-neighbor hopping terms.  The circles
represent the bottom layer, or layer 1, while the squares represent
the top layer, or layer 2.  The intralayer hopping is $t$, while the
interlayer hopping is $t_\bot$.  We label the dimerized sites as $a_1$
and $a_2$, while the non-dimerized sites are labeled $b_1$ and $b_2$.
$(b)$ Illustration of the (reciprocal) $\bk$ space, with the parabolic
degeneracy points $\bK=\frac{4\pi}{3\sqrt{3}a}\hat{\bx}$ and $\bK'=-\bK$,
where $a$ is the (real-space) lattice constant, marked.}
\end{figure}
Our starting point will be a tight-binding A-B stacked honeycomb bilayer
lattice with only nearest-neighbor hopping terms and a two-particle
interaction.  The lattice and the corresponding reciprocal momentum
($\bk$) space are illustrated in Figure \ref{HCBilayer_Diagram}.  By
including only nearest-neighbor hopping, we are fine-tuning our system
so that it only possesses two parabolic degeneracy points, rather than
four Dirac cones.  As stated earlier, this will make our system unstable
to symmetry-breaking phases for arbitrarily weak interactions, thus
justifying the use of perturbative RG methods.  The Hamiltonian for
this system is
\begin{equation}
H=H_0^{\|}+H_0^{\bot}+H_I^{\|}+H_I^{\bot},
\end{equation}
where the kinetic energy terms\cite{CastroNetoRMP2009} are
\begin{equation}
H_0^{\|}=-t\sum_{\bR,\delta}\sum_{\sigma=\uparrow,\downarrow}[b_{1\sigma}^{\dag}(\bR+\delta)a_{1\sigma}(\bR)+b_{2\sigma}^{\dag}(\bR-\delta)a_{2\sigma}(\bR)+\text{h.c.}]
\end{equation}
and
\begin{equation}
H_0^{\bot}=-t_{\bot}\sum_{\bR}\sum_{\sigma=\uparrow,\downarrow}[a_{1\sigma}^{\dag}(\bR)a_{2\sigma}(\bR)+\text{h.c.}].
\end{equation}
Here, $t$ and $t_{\bot}$ are the intra-layer and inter-layer nearest-neighbor
hopping integrals, respectively, $\delta$ is a vector that connects an
$a_1$ site to a nearest-neighbor $b_1$ site.  The position vectors $\bR$
that we sum over are the positions of the dimerized sites.  The three
possible values of $\delta$ are $\tfrac{\sqrt{3}}{2}a\hat{\bx}+\tfrac{1}{2}a\hat{\by}$,
$-\tfrac{\sqrt{3}}{2}a\hat{\bx}+\tfrac{1}{2}a\hat{\by}$, and $-a\hat{\by}$.
Whenever there is a sum on $\delta$, we sum over all three vectors,
while, if $\delta$ occurs without a sum over it, then we choose one
such vector.

The interaction terms are given by
\begin{equation}
H_I^{\|}=\tfrac{1}{2}\sum_{k=1}^{2}\sum_{\br\br'}V_{\|}(\br-\br')[n_k(\br)-1][n_k(\br')-1]
\end{equation}
and
\begin{equation}
H_I^{\bot}=\sum_{\br\br'}V_{\bot}(\br-\br')[n_1(\br)-1][n_2(\br')-1].
\end{equation}
Here, $\br$ runs over all projections of the position vectors of the
lattice sites onto the plane of the sample, and thus is entirely in
the $xy$ plane.  $c_{k\sigma}(\br)$ is the annihilation operator for
a particle at site $\br$, and $n_k(\br)=\sum_{\sigma}c_{k\sigma}^\dag(\br)c_{k\sigma}(\br)$.  The
interaction $V(\br)$ is assumed to depend only on distance, i.e.,
$V(\br)=V(|\br|)$.  For convenience, we introduced the notation
$V_{\|}(\br)=V(\br)$ and $V_{\bot}(\br)=V(\br\pm c\hat{\bz})$, where
$c$ is the distance between the two layers.  The system represented by
this Hamiltonian will be at half filling when the chemical potential
is zero.  This follows from the fact that the Hamiltonian is invariant
under the particle-hole transformation,
\begin{eqnarray}
a_{1\sigma}(\bR)&=&\tilde{a}_{1\sigma}^{\dag}(\bR), \\
a_{2\sigma}(\bR)&=&-\tilde{a}_{2\sigma}^{\dag}(\bR), \\
b_{1\sigma}(\bR)&=&-\tilde{b}_{1\sigma}^{\dag}(\bR), \\
b_{2\sigma}(\bR)&=&\tilde{b}_{2\sigma}^{\dag}(\bR).
\end{eqnarray}
Using this, the calculated expectation value of the particle number on
a given site can be shown to be $1$.

\section{Low-energy effective theory}
Before performing our RG analysis of the above Hamiltonian, we first
derive the low-energy effective theory associated with it.  We assume that
the kinetic energy terms are dominant, and that the interactions are
small.  This will allow us to focus on the interaction-induced scattering
processes in the vicinity of $\pm\bK=\pm4\pi/(3\sqrt{3}a)\hat{\bx}$.
Note that there are four sites per unit cell, and therefore there are four bands.  For every state with
energy $E(\bk)$, there is another with energy $-E(\bk)$.  To proceed, we first write the
partition function for the system as a coherent-state path integral and
integrate out the dimerized sites $a_1$ and $a_2$.  To see that the high-
energy modes are associated with these sites, note that the separation of the
high-energy, split off, bands is set by the hopping integral
$t_\perp$ between $a_1$ and $a_2$.  We then expand around the two parabolic
degeneracy points in the Brillouin zone, associated with the remaining two bands.
The derivation of this theory is a straightforward, if tedious, generalization
of the steps followed in Ref.\ \onlinecite{VafekPRB2010}.  The low-energy
effective theory that we obtain is
\begin{equation}
Z=\int\mathcal{D}[\psi^\ast,\psi]\,\exp\left (-\int_{0}^{\beta}d\tau\,L_{\text{eff}}\right ), \label{Eq:PartFunc}
\end{equation}
where the Lagrangian $L_{\text{eff}}$ is given by
\begin{eqnarray}
L_{\text{eff}}&=&\int d^2\bR\,\psi^{\dag}\left [\frac{\partial}{\partial\tau}+H(\bp)\right ]\psi\cr
&+&\tfrac{1}{2}\sum_{S\in\mathcal{G}}g_S\int d^2\bR\,(\psi^{\dag}S\psi)^2-\mu'\int d^2\bR\,\psi^{\dag}\psi. \nonumber \\ \label{spin_1/2_eff_act}
\end{eqnarray}
Here, $\psi(\br,\tau)=[\psi_{\uparrow}(\br,\tau),\psi_{\downarrow}(\br,\tau)]^T$
is an eight-component spinor in layer (1, 2), valley ($\bK$, $-\bK$), and spin
($\uparrow$, $\downarrow$) space, and
\begin{equation}
\psi_{\sigma}(\br,\tau)=
\begin{bmatrix}
b_{1,\bK,\sigma}(\br,\tau) \\
b_{2,\bK,\sigma}(\br,\tau) \\
b_{1,-\bK,\sigma}(\br,\tau) \\
b_{2,-\bK,\sigma}(\br,\tau)
\end{bmatrix}.
\end{equation}
The matrix, $H(\bp)$, is given by
\begin{equation}
H(\bp)=\frac{p_x^2-p_y^2}{2m^*}\Sigma_x+\frac{p_xp_y}{m^*}\Sigma_y, \label{Eq:Hofp}
\end{equation}
where $m^*=2t_\perp/9a^2t^2$ is the effective mass,
$\Sigma_x=1_2\sigma^x1_2=\gamma_2\otimes 1_2$, and
$\Sigma_y=\tau^z\sigma^y1_2=\gamma_1\otimes 1_2$.  In the
former definitions, the $\tau$ matrices act in valley space,
the $\sigma$ matrices act in layer space, and the third matrix,
which is the identity in both cases, acts in spin space.  The
sum on $S$ is over all $16$ matrices of the form, $\tau_i\sigma_j1_2$.
These $16$ matrices may be classified according to the representation
of the space group under which our system is symmetric, namely the
$D_{3d}$ point group plus appropriate translations, that they transform
under.  A complete classification of the $16$ possible $4\times 4$ matrices
that act in valley and layer space is done in Ref.\ \onlinecite{CvetkovicArXiv};
we repeat those results here for convenience:
\begin{eqnarray}
A_{1g}+&:&  1_4 \nonumber\\
A_{2g}-&:&\tau^z\sigma^z\nonumber\\
E_g+&:&(1_2\sigma^x,\tau^z\sigma^y)\nonumber\\
A_{1u}-&:&\tau^z1_2 \nonumber\\
A_{2u}+&:& 1_2\sigma^z \nonumber\\
E_u-&:& (\tau^z\sigma^x,-1_2\sigma^y)\nonumber\\
A_{1\bK}+&:&\tau^x\sigma^x;\tau^y\sigma^x \nonumber\\
A_{2\bK}-&:&\tau^x\sigma^y;\tau^y\sigma^y\nonumber\\
E_{\bK}+&:&
(\tau^x1_2,-\tau^y\sigma^z;-\tau^y1_2,-\tau^x\sigma^z).\nonumber
\end{eqnarray}
The $\pm$ at the end of each representation denotes how the
associated matrices transform under time reversal.  The coupling
constant $g_S$ is that associated with the representation that
the matrix $S$ belongs to.  Since the sum includes all $16$
matrices, there are nine coupling constants in all.  It can be
shown\cite{VafekPRB2010} that there are only nine independent
coupling constants due to the symmetries of the underlying
lattice and Fierz identities.  In the notation of Ref.\ \onlinecite{VafekPRB2010},
$g_{A_{1g}}=g_{A_1}^{(c)}$, $g_{A_{2g}}=g_{D_2}^{(c)}$, $g_{E_g}=g_{A_2}^{(c)}$,
$g_{A_{1u}}=g_{B_2}^{(c)}$, $g_{A_{2u}}=g_{C_1}^{(c)}$, $g_{E_u}=g_{B_1}^{(c)}$,
$g_{A_{1K}}=g_\alpha^{(c)}$,  $g_{A_{2K}}=g_\gamma^{(c)}$, and
$g_{E_K}=g_\beta^{(c)}$.

Provided that we start with density-density
interactions only, as we do in this case, the only (initial) non-zero
coupling constants in this theory are
\begin{eqnarray}
g_{A_{1g}}&=&\tfrac{1}{2}(V_{\|,0}+V_{\bot,N})A_{uc}, \label{Eq:gA1g_V} \\
g_{A_{2u}}&=&\tfrac{1}{2}(V_{\|,0}-V_{\bot,N})A_{uc}, \\
g_{E_K}&=&\tfrac{1}{4}V_{\|,2K}A_{uc}, \label{Eq:gEK_V}
\end{eqnarray}
where
\begin{eqnarray}
V_{\|,0}&=&\sum_{\bR}V_{\|}(\bR), \label{V_def_1} \\
V_{\bot,N}&=&\tfrac{1}{3}\sum_{\bR,\delta}V_{\bot}(\bR-\delta), \label{V_def_2} \\
V_{\|,2K}&=&\sum_{\bR}V_{\|}(\bR)\cos(2\bK\cdot\bR), \label{V_def_3}
\end{eqnarray}
We note that $g_{A_{1g}}$ and $g_{A_{2u}}$ depend only on $\bq=0$ Fourier
components of the interaction, and thus we may say that they
give us a measure of the strength of the forward scattering
induced by said interaction.  Likewise, we note that $g_{E_K}$
only depends on the $\bq=\pm2\bK$ Fourier components, and thus
it gives us a measure of the strength of the back
scattering.  Furthermore, we see that $g_{A_{2u}}$ depends on the
difference between an intra-layer interaction and an inter-layer
interaction, and thus it may be seen as a measure of the imbalance
between these two interactions.

Note that our theory includes a quadratic, chemical potential-like,
term.  We may think of the undetermined constant $\mu'$ as being
chosen in such a way as to cancel out the quadratic terms that are
generated from the quartic terms under RG.  We require that this
occur because we know that our original lattice model is at half
filling (that is, it possesses particle-hole symmetry), and therefore
this must be reflected in our effective low-energy theory as well.

We could, in principle, have also determined the value of $\mu'$
when we wrote down the above effective low-energy theory.  Strictly
speaking, we should not simply drop all of the modes above the
cutoff, as we did here, but rather integrate them out in a
perturbative scheme similar to what is done in an RG analysis. This,
at first order, will not change our quartic terms because it only
generates the tree-level quartic terms.  However, it will generate
both tree-level and one-loop contributions to the quadratic terms.
It would, however, be somewhat cumbersome and, given the above
particle-hole symmetry arguments, equally unnecessary, to determine
these one-loop contributions to the chemical potential.

\section{Determination of phases}
We are now ready to describe the results of our RG analysis.  We consider two forms
of the microscopic interaction, which are given by Eqs.\ \eqref{Eq:ScrCoul_OP_Intro}
and \eqref{Eq:ScrCoul_TP_Intro}.  In both cases, we determine our initial couplings
by first using Eqs.\ \eqref{Eq:gA1g_V}--\eqref{Eq:gEK_V}
to determine the ratios, $g_{A_{2u}}/g_{A_{1g}}$ and $g_{E_K}/g_{A_{1g}}$,
as a function of the range of the interaction.  We then consider different
values of $g_{A_{1g}}$ to multiply these ratios by; we may therefore
think of this value of $g_{A_{1g}}$ as determining the overall strength
of the interaction.  We then use these couplings as the initial conditions
for the RG equations derived in Ref.\ \onlinecite{CvetkovicArXiv}, which
we integrate numerically.  We then adjust the temperature until we see
the couplings diverge as we take the scale parameter to infinity, i.e.,
until we reach the critical temperature.  We monitor the susceptibilities
to different phases as we lower the temperature, and consider a phase to
be present below the critical temperature if the corresponding susceptibility
diverges as we approach the critical temperature.

\subsection{Screened Coulomb-like interaction; two-plate case}
The first interaction form that we will consider is a screened Coulomb-like
interaction, with the screening being due to two infinite planar conducting
plates, between which the charge is located.  We consider this case because
of experiments with gated bilayer graphene, such as the experiments in Refs.\ \onlinecite{YacobyScience2010}
and \onlinecite{VelascoNatNano2012}; in these cases, the gates will serve as
the conducting plates.  We will assume that the distance between the plates
is much larger than the distance between the two layers so that we may assume
that the particles are exactly halfway between the two plates.  If the distance
between the plates is $\xi$, then the interaction is given by
\begin{equation}
V(\br)=U_0 \sum_{n=-\infty}^{\infty}\frac{(-1)^n}{\sqrt{(r/\xi)^2+n^2}}. \label{int_screened_c}
\end{equation}
As shown in Appendix \ref{App:LDB_ScrCoul}, for $|\br|\gg \xi$, we may approximate
the sum as
\begin{equation}
V(\br)\approx U_0\frac{2\sqrt{2}e^{-\pi|\br|/\xi}}{\sqrt{|\br|/\xi}}. \label{int_scr_c_long_range}
\end{equation}
The above form is useful in practice when evaluating the values of the initial
coupling constants.  In the opposite limit, $|\br|\ll \xi$, we may simply approximate
the interaction with the first few terms of the sum around $n=0$.

Note that, as is, the on-site interaction given by our formula is infinite,
and thus it would give us infinite values for the initial coupling constants.
However, we recognize that, for two electrons on the lattice that are sufficiently
close, the electrons are not localized at a single point, but rather their wave
functions have a finite extent in space.  This will render the on-site interaction
finite.  As a simple model of this effect, we set the on-site interaction equal
to some constant $\lambda$ times the nearest-neighbor interaction.  In our
calculations, we set $\lambda=1.2$.

\begin{figure}[tb]
\centering
\includegraphics[width=\columnwidth,clip]{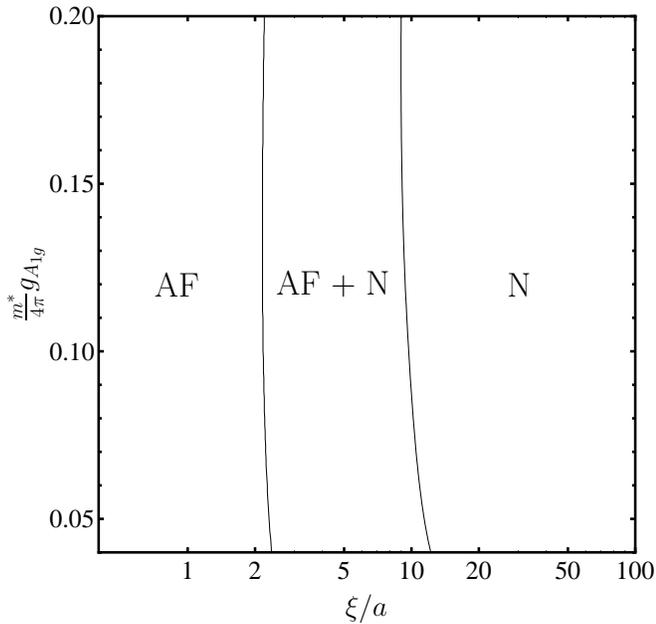}
\caption{\label{Fig:PD_TwoPlate}Phase diagram for the two-plate screened Coulomb-like interaction, Eq.\ \eqref{int_screened_c}, as a function of the dimensionless coupling strength $\frac{m^*}{4\pi}g_{A_{1g}}$ and the interaction range $\xi$ in units of the lattice spacing $a$.  For very short ranges, we find an antiferromagnetic (AF) phase.  As we increase the range, we enter a region where the susceptibilities toward both the AF and nematic (N) phases diverge as we lower the temperature.  Determining whether or not these two phases truly coexist requires a theory valid below the critical temperature, the development of which is beyond the scope of the present work.  Further increasing the range brings the system into a pure nematic state.  Note that the critical range for each of these transitions is weakly dependent on the coupling strength, and corresponds to effectively turning off back scattering.}
\end{figure}

The resulting phase diagram is shown in Figure \ref{Fig:PD_TwoPlate}.  We see
that, for very short ranges of the interaction $\xi$, the system is in an
antiferromagnetic state, in agreement with the zero-temperature results
obtained in the previous work\cite{VafekPRB2010}.  As we increase the range,
we will enter a region in which both the AF and nematic susceptibilities
diverge\cite{CvetkovicArXiv}.  This happens when $\xi$ is larger than about
two lattice spacings.  This indicates a possible coexistence of the two phases.
However, to determine if such a coexistence is in fact present would require
a theory valid below the critical temperature.  The construction of such a
theory is beyond the scope of the present work.  If we increase the range
even further, we enter a pure nematic state, again in agreement with the
previous work\cite{VafekYangPRB2010}.  This happens when $\xi$ exceeds about
$10$ lattice spacings.  Note that there is a weak dependence of these critical
ranges on the initial value of $g_{A_{1g}}$.

\subsection{Screened Coulomb-like interaction; one-plate case (dipole-like interaction)}
\begin{figure}[tb]
\centering
\includegraphics[width=\columnwidth,clip]{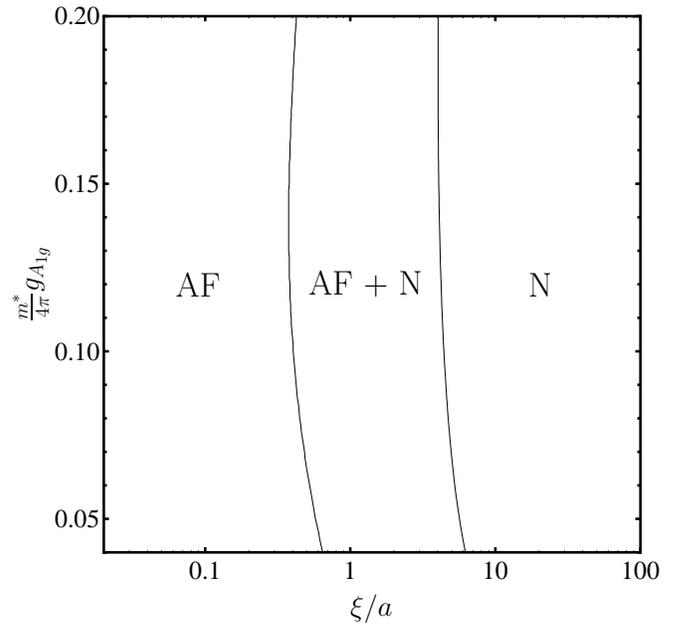}
\caption{\label{Fig:PD_OnePlate}Phase diagram for the one-plate screened Coulomb-like interaction, Eq.\ \eqref{dipole-like_int}, as a function of the coupling strength $g_{A_{1g}}$ and the interaction range $\xi$.  The phase diagram is qualitatively identical to that obtained for the two-plate case, except that the critical ranges are smaller.}
\end{figure}

The second form of the interaction that we consider is
a dipole-like interaction much like the one produced by an electron in the
presence of a single infinite conducting plate.  This interaction has the form,
\begin{equation}
V(\br)=U_0\left [\frac{1}{r/\xi}-\frac{1}{\sqrt{(r/\xi)^2+1}}\right ], \label{dipole-like_int}
\end{equation}
This interaction has a longer range than in the two-plate case, since this
falls off as $r^{-3}$ for long distances, rather than as an exponential.  As
in the previous case, this formula, as is, will give us an infinite on-site
interaction.  We use the same method as before to render this interaction finite,
and we again set $\lambda=1.2$.  The resulting phase diagram is shown in Figure \ref{Fig:PD_OnePlate}.
We note that it is qualitatively identical to that obtained from the previous
case, except that the critical ranges are smaller.  We enter the AF + nematic
``coexistence'' region when $\xi$ exceeds a value between $0.4$ and $0.6$ lattice
spacings and the pure nematic state when $\xi$ exceeds a value between $4$ and $6$
lattice spacings, depending on the initial $g_{A_{1g}}$.

Note that, throughout this section, we have been working with monotonically-decreasing
repulsive density-density interactions, and thus we only observe two of the
possible phases that we may find in the system.  For short-range interactions,
we start in the upper-right-hand corner of Fig.\ 5 of Ref.\ \onlinecite{CvetkovicArXiv},
corresponds to the AF phase.  As we increase the range, we move toward the center
of the diagram, while staying within the upper-right quadrant, passing through
the AF + nematic ``coexistence'' region and ending in the pure nematic region.
While, for $\lambda=1.2$ we only find the AF and nematic phases for the one- and two-
plate cases that we considered, there are more possible phases, even for density-
density interactions.  For example, for $\lambda<1$, such that the on-site repulsion
is {\it weaker} than the nearest-neighbor repulsion, and thus the repulsive interaction is
non-monotonic in real space, the initial value of $g_{E_K}$ may become negative.  Under
such conditions, we will find that the susceptibility toward the quantum spin Hall phase
will diverge, though never along with the antiferromagnetic susceptibility.  This
is illustrated in Fig.\ 5 of Ref.\ \onlinecite{CvetkovicArXiv}.

\section{Analysis of the antiferromagnetic state in an applied magnetic field}
\subsection{Field dependence of the AF order parameter}
We now turn our attention to the effects of a magnetic field on the antiferromagnetic
state of the system at zero temperature.  This investigation is motivated by the
fact that a gap is observed in some experiments\cite{YacobyPRL2010,VelascoNatNano2012}
and by the fact that we predict such a phase for short-range interactions within
RG.  We will investigate these effects within the framework of variational mean
field theory.  We employ this method, rather than RG, because, in the presence of
a perpendicular magnetic field, the non-interacting energy spectrum for our problem
is discrete, rather than continuous, and thus an RG analysis of the type employed
above would be more difficult.  We start by writing down a Hamiltonian corresponding
to our effective low-energy field theory and introducing the orbital effects of the
magnetic field via minimal substitution.  We will neglect the Zeeman effect in this
case, since the spin splitting $(m^*/m_e)g\mu_BB\approx (m^*/m_e)\omega_c$ is small compared to the orbital splitting $\approx \omega_c$.
 This Hamiltonian
is
\begin{eqnarray}
H&=&\int d^2\br\,\psi^{\dag}(\br)\left (\sum_{a=x,y}d^a(\piv)\Sigma^a\right )\psi(\br) \cr
&+&\tfrac{1}{2}\sum_{S}\int d^2\br\, g_S[\psi^{\dag}(\br)S\psi(\br)]^2, \label{Eq:H_FiniteB}
\end{eqnarray}
where $\piv = \bp-\frac{e}{c}\bA$, $\bA$ is the applied magnetic vector potential,
\begin{equation}
d^x(\piv)=\frac{\pi_x^2-\pi_y^2}{2m^*},\,d^y(\piv)=\frac{\pi_x\pi_y+\pi_y\pi_x}{2m^*},
\end{equation}
and the sum on $S$ stands for the four-fermion interaction terms that appear
in Equation \eqref{spin_1/2_eff_act}.  Here, $\psi$ is a field operator, not a Grassman
field as it was in previous sections.

Our variational mean field calculation will proceed as follows.  We start by adding
and subtracting a source term for the antiferromagnetic order parameter,
\begin{equation}
\Delta\int d^2\br\,\psi^{\dag}(\br)1_2\sigma^zs^z\psi(\br),
\end{equation}
in the Hamiltonian.  We now define two parts to the Hamiltonian, a ``non-interacting''
part, $H_0$,
\begin{eqnarray}
H_0&=&\int d^2\br\,\psi^{\dag}(\br)\left (\sum_{a=x,y}d^a(\piv)\Sigma^a\right )\psi(\br)+ \cr
&+&\Delta\int d^2\br\,\psi^{\dag}(\br)1_2\sigma^zs^z\psi(\br),
\end{eqnarray}
and an ``interaction'', $H_I$,
\begin{eqnarray}
H_I&=&\tfrac{1}{2}\sum_{S,T}\int d^2\br\, g_S[\psi^{\dag}(\br)S\psi(\br)]^2- \cr
&-&\Delta\int d^2\br\,\psi^{\dag}(\br)1_2\sigma^zs^z\psi(\br).
\end{eqnarray}
We then exactly diagonalize $H_0$, find the expectation value of the full Hamiltonian
with respect to the ground state of $H_0$, and minimize the result with respect to
$\Delta$.  We will provide the details of the diagonalization in Appendix \ref{App:Soln_H_AF},
and simply quote the main result here.  In the Landau gauge, our states can be described
in terms of a wave number $k$, an orbital index $n$, a valley index $\tau$, and a spin
index $s$; the corresponding energy eigenvalues are $E_n=\pm\sqrt{n(n-1)\omega_c^2+\Delta^2}$,
where $\omega_c=\frac{eB}{m^*c}$ is the cyclotron frequency of the electrons.  For
$n\geq 2$, each of these energy levels is four-fold degenerate due to valley and spin
degeneracies.  The levels given by $n=0$ and $1$ are also four-fold degenerate, but
this time due to \textit{orbital} ($n=0$ or $1$) and spin degeneracies.

We may now rewrite the field operators in terms of these eigenstates:
\begin{equation}
\psi(x,y)=\sum_{k,n,\tau,s}[\psi_{k,n,\tau,s}^{+}(x,y)a_{k,n,\tau,s}+\psi_{k,n,\tau,s}^{-}(x,y)b_{k,n,\tau,s}^{\dag}], \label{field_op_es}
\end{equation}
where $\psi_{k,n,\tau,s}^{+(-)}$ represents the positive (negative) energy state for
a given wave number and orbital, valley, and spin indices.  We have chosen our operators
$a$ and $b$ such that they annihilate the ground state $\left |0\right >$.  We may now
take the expectation value of our Hamiltonian with respect to the ground state of $H_0$
and minimize the result with respect to $\Delta$; we will provide some details on how
this was done in Appendix \ref{App:VarMF_AF}.  Upon doing so, we find that the equation
for $\Delta$ is, assuming $\Delta>0$,
\begin{equation}
g_{\text{eff}}\omega_c\left (\sum_{n=2}^{N}\frac{\Delta}{E_n}+1\right )=\Delta, \label{min_cond}
\end{equation}
where
\begin{equation}
g_{\text{eff}}=\frac{m^*}{4\pi}(g_{A_{1g}}+g_{A_{2u}}+4g_{E_K})
\end{equation}
and $N$ is an upper cutoff on the orbital index, which we impose because our theory only
works for low energies and because, in reality, we do not have electronic states in our
system at arbitrarily large energies.  In the limit of zero magnetic field, we may treat
the sum as a Riemann sum, with $\epsilon=n\omega_c$ and $\delta\epsilon=\omega_c$; our
equation then reduces to
\begin{equation}
g_{\text{eff}}\omega_c\int_{0}^{\Omega}\frac{d\epsilon}{\sqrt{\epsilon^2+\Delta_0^2}}=1, \label{min_cond_zero_field}
\end{equation}
where $\Omega$ is an upper cutoff on the energy and we introduce $\Delta_0$ as the value
of the AF order parameter in the absence of an applied magnetic field; $N$ is related to
the energy cutoff by $\Omega=\omega_c\sqrt{N(N-1)}$.  We have verified this result with a
separate calculation.  As is, we cannot send the upper cutoff to infinity in our equation
without encountering a divergence in the sum, since the summand only decreases as $\frac{1}{n}$.
We can, however, rewrite the equation in such a way that we can do this if we eliminate
$g_{\text{eff}}$ in favor of $\Delta_0$.  We give the details on how we do so in Appendix \ref{App:VarMF_AF};
the final result is
\begin{equation}
I(\alpha)+\frac{1}{\alpha}-\ln\left (1+\sqrt{\alpha^2+\tfrac{3}{4}}\right )=\ln{\beta}, \label{AF_SC_Equation_Alt}
\end{equation}
where $\alpha=\Delta(B)/\omega_c$, $\beta=\omega_c/\Delta_0$, $\Delta(B)$ is the order
parameter at finite field, and $\omega_c$ is the cyclotron frequency of the electrons
in our system.  The function, $I(\alpha)$, the exact form of which we state in Appendix \ref{App:VarMF_AF},
is monotonically decreasing, falls off as $\alpha^{-3}$ for large $\alpha$, and may be
very accurately approximated for {\it any} value of $\alpha$ by
\begin{equation}
I(\alpha)\approx -\frac{2}{[(-\tfrac{1}{2}I(0))^{-2/3}+4\cdot 6^{2/3}\alpha^2]^{3/2}},
\end{equation}
where $I(0)\approx -0.0503$ is the exact value of $I(\alpha)$ at $\alpha=0$.

While the above equation must, in general, be solved numerically, we may obtain analytic
expressions in two limiting cases, namely the large and small $\beta$ limits (equivalently,
for large and small applied magnetic fields).  Let us first consider the large $\beta$
limit.  This means that the right-hand side of our equation is large and positive, and
thus, as implied by our above discussion, $\alpha$ should be small.  In this limit, we
may set the first and third terms on the left to their values at $\alpha=0$, since both
are finite at this point, while the second term diverges.  Our equation becomes
\begin{equation}
\frac{1}{\alpha}-C=\ln{\beta},
\end{equation}
where $C$ is the value of the first and third terms at $\alpha=0$; its value is
approximately $0.67$.  Here, we are assuming that $\alpha$ is positive.  Solving
for $\alpha$, we find that
\begin{equation*}
\alpha=\frac{1}{\ln{\beta}+C},
\end{equation*}
or
\begin{equation}
\Delta=\frac{\omega_c}{\ln(\omega_c/\Delta_0)+C}.
\end{equation}
We see that the behavior is almost linear in the magnetic field, but with a
logarithmic correction.

Next, we consider the small $\beta$ limit.  In this case, the right-hand side
of Eq.\ \eqref{min_cond} becomes large and negative, and thus $\alpha$ should become
large.  We would be tempted to drop all but the third term, since the first
two terms go to zero for large values of $\alpha$.  This would give us a result
that is only accurate at constant order in $\beta$, however.  This is because,
if we expand the third term in powers of $\alpha^{-1}$, then the next-lowest order
term after the logarithmic term is of order $\alpha^{-1}$, and the second term
in our equation is also of this order; in fact, it cancels this term exactly.
We may still drop the first term, since, as stated above, the lowest-order term
that it contributes is of order $\alpha^{-3}$.  In this case, our equation becomes
\begin{equation}
\frac{1}{\alpha}-\ln\left (1+\sqrt{\alpha^2+\tfrac{3}{4}}\right )=\ln{\beta}.
\end{equation}
Again, we are assuming that $\alpha>0$.  If we take the exponential of both sides,
we get
\begin{equation*}
e^{-1/\alpha}\left (1+\sqrt{\alpha^2+\tfrac{3}{4}}\right )=\frac{1}{\beta}.
\end{equation*}
We will now expand the left-hand side in powers of $\alpha^{-1}$ to the order
$\alpha^{-2}$.  Doing so, but first pulling out a factor of $\alpha$ from the
second factor, we get
\begin{equation*}
\alpha\left (1-\frac{1}{8\alpha^2}\right )=\frac{1}{\beta}.
\end{equation*}
If we rearrange this, we finally arrive at the quadratic equation,
\begin{equation*}
8\beta\alpha^2-8\alpha-\beta=0.
\end{equation*}
If we solve this equation and take the positive solution, we get
\begin{equation*}
\alpha=\frac{1+\sqrt{1+\tfrac{1}{2}\beta^2}}{2\beta}\approx \frac{1}{\beta}+\frac{\beta}{8}.
\end{equation*}
Rewriting this result in terms of $\Delta$ and $\omega_c$, we get
\begin{equation}
\Delta=\Delta_0+\frac{\omega_c^2}{8\Delta_0}.
\end{equation}
We see that, for low fields, the antiferromagnetic order parameter increases
quadratically with the field.

\begin{figure}[tbh]
\centering
\includegraphics[width=\columnwidth,clip]{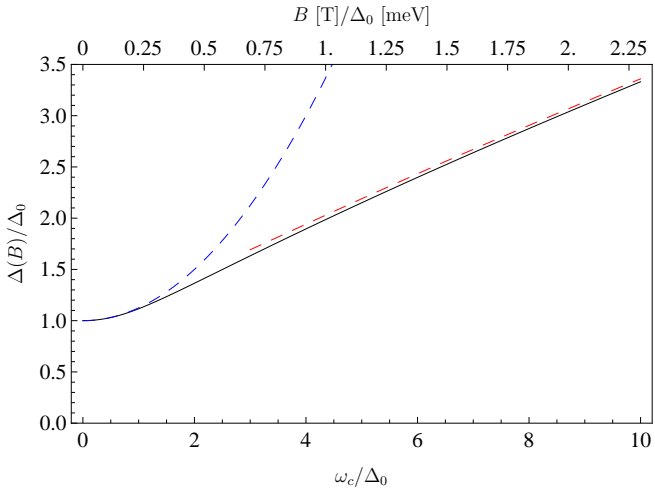}
\caption{\label{AF_SC_Solutions}Plot of the solution to Equation \eqref{AF_SC_Equation_Alt}.  The solid line is the numerical solution, while the dashed lines are the solutions in the low- and high-field limits.  The vertical axis is the value of the order parameter $\Delta$ divided by its zero-field value $\Delta_0$.  The bottom horizontal axis is the cyclotron frequency $\omega_c=eB/m^*c$ divided by $\Delta_0$, while the top horizontal axis is the applied magnetic field $B$ divided by $\Delta_0$; in determining the latter from the former, we assumed that the effective mass\cite{Mayorov2011} $m^*=0.028m_e$, and\cite{VelascoNatNano2012} $\Delta_0=0.95\text{ meV}$.}
\end{figure}

We now solve Equation \eqref{AF_SC_Equation_Alt} numerically; the numerical
result, along with the low- and high-field limits derived above, is plotted
in Figure \ref{AF_SC_Solutions}.
If we look at our low- and high-field expressions, we see that the slope of
our low-field approximation increases with $B$, while our high-field
approximation has a decreasing slope.  This implies that there should be a
maximum slope to the exact curve.  We determined the maximum slope of the
$\Delta(B)/\Delta_0$ versus $\omega_c/\Delta_0$ curve, and found that it
is about $0.2681$, and occurs when $\omega_c/\Delta_0\approx 2.432$.  These
values are independent of the values of $m^*$ and $\Delta_0$.  Using the
experimentally-determined value\cite{Mayorov2011} of the effective
mass, $m^*=0.028m_e$, and the experimentally-determined value of the order
parameter at zero field\cite{VelascoNatNano2012}, $\Delta_0=0.95\text{ meV}$, we may determine the maximum
slope of the $\Delta(B)$ versus $B$ curve.  We find that the slope is $1.11\,\frac{\text{meV}}{\text{T}}$,
and that it occurs at a field of about $0.56\text{ T}$.

\subsection{Comparison with experimental results}
We would now like to compare our theoretical results to the experimental
data\cite{VelascoNatNano2012}.  First of all, we note that $\Delta$ is
not the energy gap in our system.  In fact, the energy eigenvalues stated
earlier are an ``auxiliary spectrum'', and do not represent the true
(many-body) energy spectrum of our system.  As an approximation to the
actual energy gap, we will consider particle-hole excitations of the
``vacuum'', or trial ground state, for our system.  We construct a state,
$a^{\dag}_{\alpha}b^{\dag}_{\beta}\left |0\right >$, where $\alpha$ and
$\beta$ stand for the full sets of quantum numbers describing the particle
and hole states, respectively, and find the difference between the expectation
value of our Hamiltonian for this state and that for the trial ground state.
For both states, we assume the value of $\Delta$ that is obtained from the
minimization condition, Eq.\ \eqref{AF_SC_Equation_Alt}.  The states $\alpha$
and $\beta$ that result in the lowest value of the excitation energy will be
taken to give the actual energy gap.  We will provide details of the derivation
in Appendix \ref{App:VarMF_Ex}, and simply quote the final result here.  If we
take the electron (hole) state $\alpha$ ($\beta$) to have an orbital quantum
number $n_1$ ($n_2$), wave number $q_1$ ($q_2$), valley index $\tau_1$ ($\tau_2$),
and spin $s_1$ ($s_2$), then the excitation energy is
\begin{equation}
E_{\text{ex}}=E_{n_1}+E_{n_2}+\frac{m^*}{4\pi}g^*\omega_c(\tau_1 s_1-\tau_2 s_2), \label{Eq:EXSpectrum}
\end{equation}
where $g^*=g_{A_{1g}}+g_{A_{2u}}-4g_{E_K}$.  Here, $\tau_k=\pm 1$ if the
state exists in the $\pm\bK$ valley, and $s_k=+1$ ($-1$) for a spin up
(down) state.  The single-particle energies are $E_n=\sqrt{n(n-1)\omega_c^2+\Delta^2}$.
Note that this energy does not depend on the wave numbers of the states,
and only depends on the valley and spin indices via their products.  If
one of the states is one of the lowest Landau levels, given by $n_k=0$
or $1$, then these products are locked to a specific value.  To be exact,
if $n_1=0$ or $1$, then $\tau_1 s_1=1$.  Similarly, if $n_2=0$ or $1$,
then $\tau_2 s_2=-1$.  We see that this energy includes a term linear in
the magnetic field, and is in agreement with the results obtained in Ref.\ \onlinecite{KharitonovArXiv}.
The key difference between our derivation and that presented in Ref.\ \onlinecite{KharitonovArXiv}
is that we did not need to assume the presence of another ``order parameter''
(in fact, as we will explain shortly, this other paramter is not really an order
parameter in the sense that it breaks any additional symmetries), corresponding
to the matrix $\tau_z 1_2 s_z$ in the notation of the present paper, or a
``staggered spin current'' state\cite{CvetkovicArXiv} to obtain this linear
term, assuming that we properly calculate the excitation energy (i.e., we
calculate it from the full Hamiltonian rather than assume that the gap in
the single-particle ``auxiliary spectrum'' is the observed gap).  In fact,
the above result would not have changed had we had included this parameter
in our variational analysis, assuming that it was sufficiently small --- it
would have only introduced a constant shift to the energies in the ``auxiliary
spectrum'' and left the associated wave functions unchanged.

\begin{figure}[tbh]
\centering
\includegraphics[width=\columnwidth,clip]{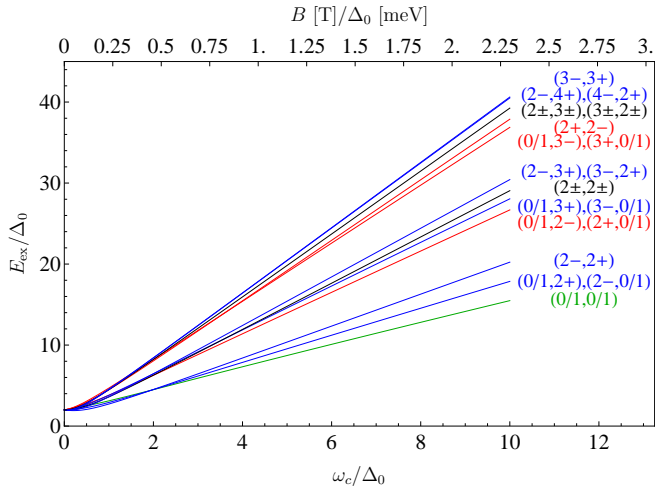}
\caption{\label{Fig:EXSpecPlot}Plot of the excitation spectrum, Eq.\ \eqref{Eq:EXSpectrum}.  The vertical axis is the excitation energy in units of $\Delta_0$, and the horizontal axes are the same as in Fig.\ \ref{AF_SC_Solutions}.  Each curve is labeled according to which electron and hole states are occupied, with a label of the form (electron, hole), with the orbital index $n$ and the sign of the product of the valley and spin indices, $\tau s$, indicated.}
\end{figure}

We present a plot of part of this excitation spectrum in Fig.\ \ref{Fig:EXSpecPlot}.
We find that the gap for low fields is, in fact, not given by taking $n_1$ and
$n_2$ to both be either $0$ or $1$, which would correspond to the lowest-energy
states in the ``auxiliary spectrum''.  Instead, it is given by taking $n_1=n_2=2$,
with $\tau_1 s_1=-1$ and $\tau_2 s_2=1$.  At higher fields, however, we find that
excitations with $n_1$ and $n_2$ both equal to either $0$ or $1$ do, in fact, give
us the actual gap.  To obtain the value for $g^*$, we fit the slope of our predicted
high-field gap at around $2.5\text{ T}$ to the slope found in Ref.\ \onlinecite{VelascoNatNano2012}
of $5.5\,\frac{\text{meV}}{\text{T}}$.  Assuming that the effective mass is given by
the experimental value\cite{Mayorov2011} of $m^*=0.028m_e$, where $m_e$ is the mass
of an electron, we obtain $\frac{m^*}{4\pi}g^*=0.44$.  Note that this differs slightly from
the value used in Ref.\ \onlinecite{KharitonovArXiv}, namely $\frac{m^*}{4\pi}g^*=0.4$;
this is the value that we would obtain if we instead fit the slope of the high-field
gap at the point where the AF order parameter reaches its maximum slope to the
experimental value.  For the value of $g_{\text{eff}}$ that we use, we find that the
gap has a minimum at a non-zero value of the field; the minimum is reached at a field
of $B\approx 0.047\text{ T}$, and is $E_g\approx 1.91\Delta_0$.  We also find a kink
in the field dependence of the gap, at which the gap goes from being given by
$n_1=n_2=2$, $\tau_1 s_1=-1$, and $\tau_2 s_2=1$ to being given by $n_1$ and $n_2$
either $0$ or $1$.  This kink appears at a field of $B\approx 0.45\text{ T}$.

We also considered the data from Ref.\ \onlinecite{YacobyPRL2010}.  In this
case, we fit our expression to the $\nu=0$ gap presented therein.  Note that,
from the low-field data, it is unclear what the size of the energy gap, if any,
is.  Nevertheless, we can still obtain a fit to the slope of the gap at high fields,
since the zero-field value of the order parameter only enters via a small logarithmic
correction to the high-field slope.  The experimentally-determined slope is
$1.7\,\frac{\text{meV}}{\text{T}}$.  If we perform our fit in the same way
as before, we obtain $\frac{m^*}{4\pi}g^*=-0.018$.  In this case, since we
obtain a negative value for $g^*$, we will find that energy of the $n_1$
and $n_2$ equal to $0$ or $1$ excitation has a non-monotonic dependence on
the magnetic field.  In fact, it gives us the energy gap for all
fields.  It possesses a very shallow minimum of $1.99985\Delta_0$ at a
field of about $0.017\text{ T}$ and, unlike the previous case, there is
no kink.  In this case, we cannot completely rule out the possibility
of the gap actually possessing such a minimum on the basis of the data
given in Ref.\ \onlinecite{YacobyPRL2010} alone, due to the lack of
data at low fields.  Note that we required a negative value of $g^*$,
which would imply that $4g_{E_K}>g_{A_{1g}}+g_{A_{2u}}$, to fit the data.
Satisfying this inequality would require either an attractive interaction
or one that is non-monotonic; this may be seen by noting that it is equivalent
to $V_{\|,2K}>V_{\|,0}$, which cannot be satisfied for any monotonically-decreasing
repulsive interaction.

Note that, while we are able to fit the experimental data\cite{YacobyPRL2010,VelascoNatNano2012}
at high fields, we also predict finer features at low fields that are
not resolved in these experiments, namely a slight non-monotonic behavior
of the gap and, in the case of our fit to the data from Ref.\ \onlinecite{VelascoNatNano2012},
a ``kink''.  It is possible that such features are, in fact, present,
but cannot be observed in the experiments due, for example, to the fact
that, at finite temperature, any sharp features that would have appeared
at zero temperature are ``washed out'', thus introducing uncertainty into
any energy gaps extracted from the data.  It is also possible that these
features, which are predicted from a mean-field calculation, will be
removed once fluctuations are taken into account.  The development of a
more sophisticated method for treating this problem is therefore of
interest, but it is beyond the scope of the present paper.

\subsection{Symmetry analysis in the presence of an applied magnetic field}
In the absence of a magnetic field, the honeycomb bilayer lattice considered
in this paper possesses a $D_{3d}$ point group symmetry.  The AF and ``staggered
spin current'' orders transform under different representations of this
group --- the AF state transforms under the $A_{2u}$ representation, while
the ``staggered spin current'' state transforms under the $A_{1u}$
representation\cite{CvetkovicArXiv}.  Physically, this means that the AF
order parameter is even under mirror reflections and odd under $C'_2$
rotations, while the ``staggered spin current'' order parameter is odd
under mirror reflections and even under $C'_2$ rotations.  Note that both
orders are odd under inversion.  In fact, the two also transform differently
under time reversal; the AF order parameter is odd, while the ``staggered
spin current'' is even\cite{CvetkovicArXiv}.  This means that the expectation
value of the ``staggered spin current'' operator must vanish in the AF state.

When we apply a magnetic field, however, the point group is reduced to $S_6$.
This is because the magnetic field is an axial vector that is odd under mirror
reflections and even under inversion.  In this case, the AF and ``staggered
spin current'' orders transform under the same representation, namely the
$A_{1u}$ representation\cite{TinkhamBook}.  Physically, this is because the
mirror reflection and $C'_2$ symmetries are no longer present.  As pointed
out above, the two order parameters transform differently under time reversal;
however, time reversal symmetry is broken in the presence of a magnetic field.
This means that the two orders no longer break different symmetries, and thus
there is nothing preventing the system from acquiring a non-zero expectation
value of one of these order parameters in the presence of the other.

The development of a finite expectation value of the ``staggered spin current''
operator was correctly pointed out in Ref.\ \onlinecite{KharitonovArXiv}, but
was attributed to ``the emergence of the $n=0,1$ Landau levels (LLs) and the
peculiar property of their wave-functions to reside on only one sublattice in
each valley''.  Here, we show that it must be present on much more general
grounds, and is not tied to the properties of the Landau levels.

At $B=0$, the AF order parameter breaks time reversal and inversion symmetry,
but it preserves mirror reflection symmetry\cite{CvetkovicArXiv}.  Therefore,
the wave functions for this state are eigenstates of the reflection operators
as well, and may be classified as even or odd under them.  Let us now consider
the expectation value of an observable $O$ that transforms under the $A_{1u}$
representation of the $D_{3d}$ point group, such as the ``staggered spin current''
order parameter.  This operator will have the property that any mirror reflection
$\sigma_d$ will anticommute with it, i.e. $\sigma_d O=-O\sigma_d$.  Because of
this, the expectation value of the ``staggered spin current'' operator with
respect to the AF state of the Hamiltonian must be zero.

Let us now consider the case in which $B\neq 0$.  As stated before, this will
break the mirror reflection symmetry of our system.  However, it is symmetric
under such a reflection followed by a reversal of the magnetic field (i.e.,
$B\rightarrow -B$).  This means that we may classify all eigenstates as even
or odd under this combination of operations.  This means that, in terms of the
eigenstates of the Hamiltonian at $B=0$, we may write the new AF state of the
Hamiltonian in the presence of an applied field as
\begin{equation}
\left |AF(B)\right >=\sum_{i}[\alpha^{(e)}_i(B)\left |i,e\right >+\alpha^{(o)}_i(B)\left |i,o\right >],
\end{equation}
where $\left |i,e\right >$ ($\left |i,o\right >$) is a general even (odd) state
of the zero-field Hamiltonian.  One set of the $\alpha$ coefficients (i.e.,
$\alpha^{(e)}_i$ or $\alpha^{(o)}_i$) must be even functions of $B$, while the
other must be odd.  If we now calculate the expectation value of $O$ with respect
to this state, only matrix elements that mix states of opposite parity under
reflections will appear:
\begin{equation}
\left <O\right >=\sum_{i,j}(\alpha^{(e)}_i)^*\alpha^{(o)}_j\left <i,e\right |O\left |j,o\right >+\text{c.c.}
\end{equation}
Since one of either the $\alpha^{(e)}_i$ or $\alpha^{(o)}_i$ must be even
functions of $B$, while the others must be odd, we see that the expectation
value of $O$ must be an odd function of $B$.

If we calculate the expectation value of the ``staggered spin current''
operator for the trial ground state that we work with above, we find that
it is a linear function of $B$.  This is consistent with our general
conclusions and with the observation made by Kharitonov\cite{KharitonovArXiv}.

\subsection{Corrections to the ``auxiliary spectrum''}
There is one other point that we wish to address.  Throughout this
calculation, we have been assuming that the presence of the order
parameter opens up a simple gap in the spectrum, modifying the low-energy
dispersion to have the form, $E(\bk)=\pm\sqrt{[\epsilon(\bk)]^2+\Delta^2}$,
where $\epsilon(\bk)$ is the dispersion in the absence of the order
parameter.  This assumption is, in fact, not entirely true; in reality,
the dispersion acquires a ``Mexican hat'' shape.  This can be seen by
taking the continuum limit of the effective action given by Eqs.\ (19)--(21)
in Ref.\ \onlinecite{VafekPRB2010}, which will give rise to an additional
term $(v_Fq/t_{\bot})^2\partial/\partial\tau$, where $v_F=\tfrac{3}{2}at$,
in Eq.\ (24) in the same work.  In the absence of an applied magnetic field,
but in the presence of the antiferromagnetic order parameter $\Delta_0$,
the dispersion is
\begin{equation}
E(\bk)=\pm\frac{\sqrt{(k^2/2m^*)^2+\Delta^2}}{1+(v_Fk/t_\bot)^2}.
\end{equation}
If we expand this expression to fourth order in $k$, we obtain
\begin{equation}
E(\bk)\approx\pm\Delta\left [1-\left (\frac{v_Fk}{t_\bot}\right )^2+\left (\frac{v_Fk}{t_\bot}\right )^4+\tfrac{1}{2}\left (\frac{k^2}{2m^*\Delta}\right )^2\right ].
\end{equation}
We see that the quadratic term is negative, while the quartic term is
positive, thus giving this dispersion the ``Mexican hat'' shape.  Note
that it is the presence of a non-zero $\Delta_0$ that changes the scaling
dimensions of our operators and necessitates the inclusion of the
above-mentioned term for a complete description of the low-energy behavior.
For $\Delta_0=0$, which is the starting point of our RG analysis, the scaling
dimensions of our operators are the same as they were in Ref.\ \onlinecite{VafekPRB2010},
which justifies neglecting this term in our RG analysis.  Performing the
standard minimal coupling to an external vector potential, we have looked
at what effect this would have on the lowest Landau levels, and we find that
it lifts the orbital degeneracy of the $n=0$ and $1$ Landau levels.  To be
exact, these two levels take the form,
\begin{equation}
E_n=\frac{\Delta}{1+\frac{\omega_c}{t_\bot}(n+\tfrac{1}{2})}.
\end{equation}
For small magnetic fields, for which $\omega_c\ll t_\bot$, we find that the
splitting between these levels is approximately $\frac{\omega_c}{t_\bot}\Delta=aB$,
where $a\approx 0.0106\,\frac{\text{meV}}{\text{T}}$.  This is small compared
even to the Zeeman splitting, which is also linear in $B$, but with a slope of
about $0.232\,\frac{\text{meV}}{\text{T}}$.  Therefore, we may safely neglect
this effect in our calculations.

\section{Conclusion}
We have employed weak-coupling perturbative RG methods to investigate the different
phases that fermions on a honeycomb bilayer lattice with finite-range interactions
will enter.  The use of these methods is justified since we only include nearest-neighbor
hopping terms, resulting in a band structure with two quadratic degeneracy points
and therefore in a finite density of states at the Fermi level at half filling.  We
considered two forms of the interaction, a screened Coulomb-like interaction much
like the one produced by a point charge situated exactly halfway between two infinite
parallel conducting plates, as well as that produced by a point charge in the presence
of a single conducting plate.  For all cases, we determined what phase the system enters
as a function of the range of the interaction.

We found that the system, for both forms of the interaction, enters an antiferromagnetic
state for short ranges and a nematic state for long ranges, in agreement with the previous
work\cite{VafekPRB2010, VafekYangPRB2010}.  For intermediate ranges, we find that the susceptibilities
towards both the antiferromagnetic and nematic phases diverge, though not necessarily with the
same exponent\cite{CvetkovicArXiv}.  This indicates a possible coexistence of the two phases.
To determine whether the phases truly coexist, or if only one appears, would require a theory
that is valid below the critical temperature.  The development of such a theory is a problem of
great interest, but is beyond the scope of the present work.

While we find that short-ranged interactions result in an antiferromagnetic state, which is
gapped, while long-ranged interactions result in a nematic state, which is gapless, the ranges
at which we see a transition from one to another are too short to explain why the experiment in Ref.\
\onlinecite{VelascoNatNano2012} observes an energy gap in their sample, while that of Ref.\ \onlinecite{Mayorov2011}
finds evidence for a nematic state in their sample.  We note, however, that the sample studied in
Ref.\ \onlinecite{Mayorov2011} has a higher mobility than that studied in Ref.\ \onlinecite{VelascoNatNano2012},
which suggests that the degree of disorder present in the system is a major influence on the phase
that appears.  Throughout our work, we have assumed a clean sample; the incorporation of disorder
into our system may therefore be important in better explaining this apparent discrepancy in the
experimental results.

Another issue is the fact that we predict an antiferromagnetic phase, which breaks a continuous $SU(2)$ spin symmetry, 
which is impossible in two dimensions at finite temperature.  We should therefore
view the divergence of the susceptibility toward this phase as simply identifying the dominant ordering
tendency in this case, even if there is no actual symmetry breaking, as pointed out in Ref.\ \onlinecite{CvetkovicArXiv}.
We expect that if the RG could be carried out exactly, we would find no divergent susceptibilities in this case. 
It would be very interesting to find systematic extensions of the approximate RG analysis used here 
which would be powerful enough to capture the Mermin-Wagner physics.

We also considered the effects of an applied perpendicular magnetic field on the system when it
is in the antiferromagnetic phase. In this case the fluctuations effects are weaker than at $B=0$ 
since the broken continuous symmetry is the 
$U(1)$ subgroup of the full spin $SU(2)$ group. 
At $B\neq 0$ a finite temperature transition into a power-law correlated state is in fact possible.
Our variational mean field investigation was
motivated by the fact that we find an antiferromagnetic state in our RG calculations for short-ranged
interactions, as well as by experimental data\cite{YacobyPRL2010,VelascoNatNano2012} on the gap size
as a function of an applied magnetic field.  We find that the antiferromagnetic order parameter
increases quadratically with the field for low fields, then acquires a dependence of the form
$B/\ln(B/B_0)$ for large fields.  We also determined the gap by considering the energy required
to create particle-hole excitations of our variational ground state.  We find that this energy is
the sum of the energies of the particle and the hole given by the single-particle ``auxiliary spectrum'',
plus a term linear in the magnetic field.  The excitation that gave us the smallest such energy was
assumed to determine the energy gap in the system.  We found that the gap has a slight non-monotonic
behavior for low fields, followed by a quasi-linear increase at higher fields.  We also compared this
prediction to the experimental data and found that good agreement can be achieved. 

One reason for our switch to mean-field methods for treating this problem is that we have already
established via RG methods the presence of the antiferromagnetic state under certain conditions.  As
long as we are considering a case in which we know this phase to be present, and because said phase is
gapped, we expect that an expansion around the mean-field solution will be convergent, thus justifying
our use of such methods in studying the phenomenology of the phase.  Another reason for our use of mean-field
methods as opposed to the RG methods used previously in the zero-field case is the fact that the energy
spectrum for the non-interacting problem is discrete, rather than continuous, and momentum $\bk$ is not a good quantum number, 
making the use of RG methods
more difficult.  While we expect our mean-field methods to be fairly accurate, such methods are still only
approximate.  The problem of developing a more sophisticated technique for determining the AF order parameter
and the energy gap in the system is therefore of great interest.

\acknowledgments
The authors would like to acknowledge useful discussions and correspondence with M. Kharitonov, 
S. Kivelson, C. N. Lau, K. Novoselov, S. Raghu, B. Roy, D. Smirnov, and K. Yang.  
This work was supported in part by the NSF CAREER Award under Grant no. DMR-0955561.

\appendix
\section{Long-distance behavior of the screened Coulomb-like interaction} \label{App:LDB_ScrCoul}
We will now demonstrate that Equation \eqref{int_scr_c_long_range} is the correct long-range behavior of the screened Coulomb-like interaction, Equation \eqref{int_screened_c}.  We start by rewriting the sum using the identity,
\begin{equation}
\frac{1}{r}=\frac{2}{\sqrt{\pi}}\int_0^{\infty}du\,e^{-r^2 u^2},
\end{equation}
obtaining
\begin{equation}
V(\bR)=\frac{2}{\sqrt{\pi}}U_0\int_0^{\infty}du\,e^{-u^2(R/\xi)^2}\sum_{n=-\infty}^{\infty}(-1)^n e^{-n^2 u^2}.
\end{equation}
We may evaluate the sum in terms of the Jacobi theta function,
\begin{equation}
\vartheta_4(z,q)=\sum_{n=-\infty}^{\infty}(-1)^n q^{n^2} e^{2niz},
\end{equation}
to obtain
\begin{equation}
V(\bR)=\frac{2}{\sqrt{\pi}}U_0\int_0^{\infty}du\,e^{-u^2(R/\xi)^2}\vartheta_4(0,e^{-u^2}).
\end{equation}
We now use the identity\cite{WolframFunc},
\begin{eqnarray}
\vartheta_4(z,q)&=&\frac{2\sqrt{\pi}}{\sqrt{-\log{q}}}e^{(4z^2+\pi^2)/4\log{q}}\sum_{k=0}^{\infty}e^{k(k+1)\pi^2 /\log{q}}\cr
&\times&\cosh\left [\frac{(2k+1)\pi z}{\log{q}}\right ],
\end{eqnarray}
so that
\begin{equation}
V(\bR)=4U_0\sum_{k=0}^{\infty}\int_0^{\infty}du\,\frac{1}{u}e^{-u^2(R/\xi)^2}e^{-(k+1/2)^2\pi^2/u^2}.
\end{equation}
This integral can be evaluated in terms of modified Bessel functions of the second kind; the result is
\begin{equation}
V(\bR)=4U_0\sum_{k=0}^{\infty}K_0\left [(2k+1)\pi\frac{R}{\xi}\right ].
\end{equation}
For large values of $x$, the modified Bessel function $K_n(x)$ can be approximated as
\begin{equation}
K_n(x)\approx \sqrt{\frac{\pi}{2x}}e^{-x}.
\end{equation}
We see that, in the above sum, the most dominant term for $R\gg\xi$ is the $k=0$ term, since the values of successive terms decrease exponentially with increasing $k$.  Therefore, we arrive at the form quoted in the main text, Equation \eqref{int_scr_c_long_range}.

\section{Solution of the non-interacting Hamiltonian in a magnetic field with an AF order parameter} \label{App:Soln_H_AF}
We now provide the details of the diagonalization of $H_0$.  To diagonalize $H_0$, we first note that the source term is diagonal in layer, valley, and spin space, while the ``kinetic'' term is only diagonal in valley and spin space.  This allows us to split the problem into the diagonalization of four $2\times 2$ matrices; our wave functions will have a definite valley pseudospin and real spin orientation.  Let us consider the block corresponing to the $+\bK$ valley and spin up.  We must solve
\begin{equation}
\begin{bmatrix}
\Delta && \frac{(\pi_x-i\pi_y)^2}{2m^*} \\
\frac{(\pi_x+i\pi_y)^2}{2m^*} && -\Delta
\end{bmatrix}\psi(x,y)=E\psi(x,y),
\end{equation}
where $\psi(x,y)$ is a two-component spinor corresponding to the $+\bK$ valley and spin up components of the full eight-component spinor; the other six components are all zero.  We will work in the Landau gauge, in which $\bB=-By\hat{\bx}$.  For this gauge, the above becomes
\begin{eqnarray}
\begin{bmatrix}
\Delta && \frac{1}{2m^*}\left (p_x+\frac{eB}{c}y-ip_y\right )^2 \\
\frac{1}{2m^*}\left (p_x+\frac{eB}{c}y+ip_y\right )^2 && -\Delta
\end{bmatrix}\psi(x,y) \cr
=E\psi(x,y). \nonumber \\
\end{eqnarray}
Let us now assume the following form for $\psi(x,y)$:
\begin{equation}
\psi(x,y)=\frac{1}{\sqrt{L_x}}e^{ikx}\Phi_k(y),
\end{equation}
where $\Phi_k(y)$ is another two-component spinor.  Upon substitution into our equation, we obtain
\begin{eqnarray}
\begin{bmatrix}
\Delta && \frac{1}{2m^*}\left (k+\frac{eB}{c}y-ip_y\right )^2 \\
\frac{1}{2m^*}\left (k+\frac{eB}{c}y+ip_y\right )^2 && -\Delta
\end{bmatrix}\Phi_k(y) \cr
=E\Phi_k(y). \nonumber \\
\end{eqnarray}
We may now write this in terms of the operators,
\begin{equation}
a_k=\sqrt{\frac{m^*\omega_c}{2}}\left (\frac{k}{m^*\omega_c}+y+i\frac{p_y}{m^*\omega_c}\right ),
\end{equation}
where $\omega_c=\frac{eB}{m^*c}$ is the cyclotron frequency of the electrons in our system.  These operators may be verified to satisfy the commutation relation, $[a_k,a_k^\dag]=1$.  In terms of these operators, the equation becomes
\begin{equation}
\begin{bmatrix}
\Delta && \omega_c (a_k^\dag)^2 \\
\omega_c a_k^2 && -\Delta
\end{bmatrix}\Phi_k(y)=E\Phi_k(y).
\end{equation}
Let us now define normalized functions $\phi_{k,n}(y)$ such that $a_k^\dag a_k\phi_{k,n}(y)=n\phi_{k,n}(y)$; these will just be the usual harmonic oscillator-like wave functions that emerge in the solution of the free electron gas in a magnetic field.  These functions may be shown to satisfy the relations, $a_k\phi_{k,n}(y)=\sqrt{n}\phi_{k,n-1}(y)$ and $a_k^\dag\phi_{k,n}(y)=\sqrt{n+1}\phi_{k,n+1}(y)$.  We now assume the following form for $\Phi_k(y)$:
\begin{equation}
\Phi_k(y)=\begin{bmatrix}
\alpha_{k,n}\phi_{k,n}(y) \\
\beta_{k,n}\phi_{k,n-2}(y)
\end{bmatrix}
\end{equation}
We find that this form satisfies our equation, provided that $\alpha_{k,n}$ and $\beta_{k,n}$ satisfy
\begin{equation}
\begin{bmatrix}
\Delta && \omega_c\sqrt{n(n-1)} \\
\omega_c\sqrt{n(n-1)} && -\Delta
\end{bmatrix}\begin{bmatrix}
\alpha_{k,n} \\
\beta_{k,n}
\end{bmatrix}=E\begin{bmatrix}
\alpha_{k,n} \\
\beta_{k,n}
\end{bmatrix}
\end{equation}
and $|\alpha_{k,n}|^2+|\beta_{k,n}|^2=1$.  We have thus reduced the problem to solving for the eigenvalues and eigenvectors of a $2\times 2$ matrix.  The eigenvalues are
$E=\pm E_n$, where
\begin{equation}
E_n=\sqrt{n(n-1)\omega_c^2+\Delta^2},
\end{equation}
and the corresponding eigenvectors are given by
\begin{equation}
\alpha_{k,n}=\pm\frac{1}{\sqrt{2}}\sqrt{1\pm\frac{\Delta}{E_n}},\,\beta_{k,n}=\frac{1}{\sqrt{2}}\sqrt{1\mp\frac{\Delta}{E_n}}.
\end{equation}
In the $-\bK$ valley, the positions of the creation and annihilation operators $a_k^\dag$ and $a_k$ will be interchanged; in this case, we must instead assume that
\begin{equation}
\Phi_k(y)=\begin{bmatrix}
\alpha_{k,n}\phi_{k,n-2}(y) \\
\beta_{k,n}\phi_{k,n}(y)
\end{bmatrix}.
\end{equation}
This will give us the same eigenvalue problem as before.  For the spin down case, we simply reverse the sign on $\Delta$ in our equations; we obtain the same eigenvalues as before, but $\alpha_{k,n}$ and $\beta_{k,n}$ will switch values.  This implies that, at least for $n\geq 2$, each of our energy levels is four-fold degenerate, due to valley and spin degeneracies.  For $n=0$ or $1$, on the other hand, there is no valley degeneracy for a given spin.  In these cases, the eigenfunctions are
\begin{equation}
\Phi_k(y)=\begin{bmatrix}
\phi_{k,n}(y) \\
0
\end{bmatrix}
\end{equation}
in the $+\bK$ valley and
\begin{equation}
\Phi_k(y)=\begin{bmatrix}
0 \\
\phi_{k,n}(y)
\end{bmatrix}
\end{equation}
in the $-\bK$ valley.  The former corresponds to the energy eigenvalue, $E=\Delta$, while the other corresponds to $E=-\Delta$.  Each of these levels is still four-fold degenerate, but this time due to \textit{orbital} and spin degeneracies.  Note that we may still use the wave functions quoted earlier even for this case if we adopt the convention that $\phi_{k,n}(y)$ is identically zero if $n<0$.

\section{Derivation of the variational mean-field equation for an AF order parameter} \label{App:VarMF_AF}
We now describe how to derive the variational mean-field equation, Eq.\ \eqref{AF_SC_Equation_Alt}.  We first provide details on the calculation and minimization of the ground-state expectation value of the interacting Hamiltonian with an antiferromagnetic order parameter.  Throughout this derivation, we will assume that $\Delta>0$.  We start by rewriting the field operators in our Hamiltonian in terms of the eigenstates of the $H_0$ derived above; the formula for this is given by Equation \eqref{field_op_es}.  If we label the positive energy eigenvalues as $E_{k,n,\tau,s}=E_n=\sqrt{n(n-1)\omega_c^2+\Delta^2}$, then $H_0$ becomes
\begin{eqnarray}
H_0&=&\sum_{k,n,\tau,s}E_{k,n,\tau,s}(a_{k,n,\tau,s}^{\dag}a_{k,n,\tau,s}+b_{k,n,\tau,s}^{\dag}b_{k,n,\tau,s}) \cr
&-&\sum_{k,n,\tau,s}E_{k,n,\tau,s}.
\end{eqnarray}
The expectation value of $H_0$ with respect to the ground state $\left |0\right >$ is then
\begin{equation}
\left <0\right |H_0\left |0\right >=-\sum_{k,n,\tau,s}E_{k,n,\tau,s}=-4d\sum_{n}E_n-4d|\Delta|,
\end{equation}
where $d$ is the degeneracy of each Landau level due to the wave number $k$.

Now we turn our attention to the interaction, starting with the quadratic term,
\begin{equation}
H_I^{(2)}=-\Delta\int d^2\br\,\psi^{\dag}(\br)1_2\sigma^zs^z\psi(\br).
\end{equation}
In terms of the eigenstates of $H_0$, this becomes
\begin{equation}
H_I^{(2)}=-\Delta\int d^2\br\,\sum_{mn}\psi_{m}^{\dag}(\br)1_2\sigma^zs^z\psi_{n}(\br)a_{m}^{\dag}a_{n}.
\end{equation}
Here, we let the indices $m$ and $n$ stand for all of the quantum numbers characterizing a given eigenstate, and we use $a_m$ to stand for either a positive or negative energy state, with the understanding that the negative energy state is given by $a_m=b_m^\dag$.  Upon taking the expectation value of this term with respect to the ground state, we find that only the negative energy states contribute:
\begin{equation}
\left <0\right |H_I^{(2)}\left |0\right >=-\Delta\int d^2\br\,\sum_{m}(\psi_{m}^{-})^{\dag}(\br)1_2\sigma^zs^z\psi_{m}^{-}(\br)
\end{equation}
Evaluating all of the sums and integrals using the expressions given above for the wave functions, we may write this as
\begin{equation}
\left <0\right |H_I^{(2)}\left |0\right >=-L_xL_y\Delta\mbox{Tr}(1_2\sigma^zs^z\Sigma_{-}),
\end{equation}
where
\begin{eqnarray}
\Sigma_{-}&=&\sum_{m}\psi_m^{-}(\br)(\psi_m^{-})^{\dag}(\br) \cr
&=&\frac{1}{4\pi\ell_B^2}\omega_c(N1_8+\tau_z\sigma_z 1_2-\tau_z 1_2\sigma_z-Y1_2 \sigma_z s_z).
\end{eqnarray}
Evaluating the trace, we obtain
\begin{equation}
\left <0\right |H_I^{(2)}\left |0\right >=\frac{L_x L_y}{2\pi l_B^2}\Delta Y,
\end{equation}
where $l_B=\sqrt{c/eB}$ is the magnetic length, $L_x$ and $L_y$ are the dimensions of our system, and
\begin{equation}
Y=\sum_{n\geq 2}\frac{\Delta}{E_n}+1.
\end{equation}

We now consider the quartic terms.  Each of these terms, neglecting the coupling constants and integrals over position, has the form,
\begin{equation}
[\psi^{\dag}(\br)S\psi(\br)]^2, \label{Eq:QuarticTermForm}
\end{equation}
where $S$ is a matrix.  Substituting in Equation \eqref{field_op_es}, and adopting the same conventions as before, this becomes
\begin{equation}
\sum_{mnpq}[\psi_{m}^{\dag}(\br)S\psi_{n}(\br)][\psi_{p}^{\dag}(\br)S\psi_{q}(\br)]a_m^{\dag}a_n a_p^{\dag} a_q.
\end{equation}
We now take the expectation value of this expression with respect to the ground state.  This expectation value will involve the expression, $\left <0\right |a_m^{\dag}a_n a_p^{\dag} a_q\left |0\right >$.  The only way for this to be non-zero is if $m$ and $q$ are negative-energy states.  We must also require that $n$ and $p$ either be both positive-energy states or both negative-energy states.  In the former case, we obtain, using the anticommutation relations for fermions,
\begin{equation*}
\left <0\right |b_m a_n a_p^{\dag} b_q^{\dag}\left |0\right >=\delta_{mq}\delta_{np}.
\end{equation*}
In the latter case, we obtain
\begin{equation*}
\left <0\right |b_m b_n^{\dag} b_p b_q^{\dag}\left |0\right >=\delta_{mn}\delta_{pq}.
\end{equation*}
Putting these results together, the above quartic form becomes
\begin{eqnarray}
\sum_{m,n}[(\psi_{m}^{-})^{\dag}(\br)S\psi_{n}^{+}(\br)][(\psi_{n}^{+})^{\dag}(\br)S\psi_{m}^{-}(\br)] \cr
+\sum_{m,p}[(\psi_{m}^{-})^{\dag}(\br)S\psi_{m}^{-}(\br)][(\psi_{p}^{-})^{\dag}(\br)S\psi_{p}^{-}(\br)],
\end{eqnarray}
Upon evaluating all sums and integrals, this becomes
\begin{equation}
\mbox{Tr}(S\Sigma_{+}S\Sigma_{-})+[\mbox{Tr}(S\Sigma_{-})]^2,
\end{equation}
where
\begin{eqnarray}
\Sigma_{+}&=&\sum_{m}\psi_m^{+}(\br)(\psi_m^{+})^{\dag}(\br) \cr
&=&\frac{1}{4\pi\ell_B^2}\omega_c(N1_8+\tau_z\sigma_z 1_2+\tau_z 1_2\sigma_z+Y1_2 \sigma_z s_z).
\end{eqnarray}
Evaluating the traces, we find that the total energy $E_{\text{var}}=\left <0\right |H\left |0\right >$ of our system is, noting that $d=L_x L_y/(2\pi l_B^2)$,
\begin{widetext}
\begin{eqnarray}
E_{\text{var}}&=&-\frac{2L_xL_y}{\pi\ell_B^2}\left (\sum_{n\geq 2}E_n+\Delta\right )+\frac{2L_xL_y}{\pi\ell_B^2}\Delta Y+4L_xL_y(g_{A_{1g}}+g_{A_{2u}}+4g_{E_K})\left [\left (\frac{N}{4\pi\ell_B^2}\right )^2-\left (\frac{Y}{4\pi\ell_B^2}\right )^2\right ] \cr
&+&\tfrac{1}{2}L_xL_yg_{A_{1g}}\left (\frac{2N}{\pi\ell_B^2}\right )^2,
\end{eqnarray}
where
\begin{equation}
N=\sum_{n\geq 2}1+1.
\end{equation}
We will now minimize this energy with respect to $\Delta$.  First, we take the derivative of the above expression, which may be written as
\begin{equation}
\frac{\partial E_{\text{var}}}{\partial\Delta}=\frac{2L_x L_y}{\pi l_B^2}\left [\Delta-\frac{1}{4\pi l_B^2}(g_{A_{1g}}+g_{A_{2u}}+4g_{E_K})Y\right ]\sum_{n\geq 2}\frac{1}{E_n}\left (1-\frac{\Delta^2}{E_n^2}\right ).
\end{equation}
\end{widetext}
We now set this derivative to zero.  We note that the second factor can never be zero, since $E_n>\Delta$ for all $n\geq 2$, and therefore it is always positive.  We may therefore drop this factor.  The remainder, upon simplifying and noting that $l_B=1/\sqrt{m^*\omega_c}$, thus yields Eq.\ \eqref{min_cond}.

We now wish to rewrite Eq.\ \eqref{min_cond} in such a way that we may send the upper cutoff to infinity.  We start be rewriting the sum as an integral over a ``Dirac comb'':
\begin{eqnarray}
&&\omega_c\left (\int_{3/2}^{N}d\lambda\,\frac{1}{\sqrt{\lambda(\lambda-1)\omega_c^2+\Delta^2}}\sum_{n=-\infty}^{\infty}\delta(\lambda-n)+\frac{1}{\Delta}\right ) \cr
&&=\frac{1}{g_{\text{eff}}}.
\end{eqnarray}
We now use the identity,
\begin{equation}
\sum_{n=-\infty}^{\infty}\delta(\lambda-n)=\sum_{k=-\infty}^{\infty}e^{2\pi ik\lambda},
\end{equation}
to obtain
\begin{eqnarray}
\omega_c\int_{3/2}^{N}d\lambda\,\frac{1}{\sqrt{\lambda(\lambda-1)\omega_c^2+\Delta^2}}+ \cr
+2\omega_c\sum_{k=1}^{\infty}\int_{3/2}^{N}d\lambda\,\frac{\cos(2\pi k\lambda)}{\sqrt{\lambda(\lambda-1)\omega_c^2+\Delta^2}}+\frac{\omega_c}{\Delta} \cr
=\frac{1}{g_{\text{eff}}}.
\end{eqnarray}
The integral that appears in the sum converges for all $k\geq 1$, so we may already send the upper limit to infinity.  However, we must still take care of the first integral, which will diverge if we do the same with it.  We may use the zero-field equation, Equation \eqref{min_cond_zero_field}, to rewrite the right-hand side such that we eliminate $g_{\text{eff}}$ from the equation.  If we do this and also introduce the change of variables, $\epsilon^2=\lambda(\lambda-1)\omega_c^2$, into the first integral, we may then move the first integral to the right-hand side, obtaining an integral that converges if we send the upper limit to infinity, namely
\begin{eqnarray}
&&\omega_c\int_{0}^{\infty}d\epsilon\,\left (\frac{1}{\sqrt{\epsilon^2+\Delta_0^2}}-\frac{\epsilon}{\sqrt{\epsilon^2+\tfrac{1}{4}\omega_c^2}}\frac{1}{\sqrt{\epsilon^2+\Delta^2}}\right )+ \cr
&&+\omega_c\int_{0}^{\omega_c\sqrt{3}/2}d\epsilon\,\frac{\epsilon}{\sqrt{\epsilon^2+\tfrac{1}{4}\omega_c^2}}\frac{1}{\sqrt{\epsilon^2+\Delta^2}},
\end{eqnarray}
where $\Delta_0$ is the value of $\Delta$ at zero magnetic field.  Upon evaluating this integral, our equation becomes
\begin{eqnarray}
2\omega_c\sum_{k=1}^{\infty}\int_{3/2}^{\infty}d\lambda\,\frac{\cos(2\pi k\lambda)}{\sqrt{\lambda(\lambda-1)\omega_c^2+\Delta^2}}+\frac{\omega_c}{\Delta} \cr
=\ln\left (\frac{\omega_c}{\Delta_0}+\frac{\sqrt{\Delta^2+\tfrac{3}{4}\omega_c^2}}{\Delta_0}\right ).
\end{eqnarray}
This equation may be rewritten in terms of the dimensionless parameters, $\alpha=\Delta/\omega_c$ and $\beta=\omega_c/\Delta_0$:
\begin{eqnarray}
&&2\sum_{k=1}^{\infty}\int_{3/2}^{\infty}d\lambda\,\frac{\cos(2\pi k\lambda)}{\sqrt{\lambda(\lambda-1)+\alpha^2}}+\frac{1}{\alpha}-\ln\left (1+\sqrt{\alpha^2+\tfrac{3}{4}}\right ) \cr
&&=\ln{\beta}. \label{AF_SC_Equation}
\end{eqnarray}
We can see that the left-hand side is a monotonically-decreasing function of $\alpha$ for $\alpha>0$.  In fact, as $\alpha\rightarrow 0^{+}$, the left-hand side increases indefinitely due to the second term, while, as $\alpha\rightarrow \infty$, the expression decreases indefinitely due to the third term.  This equation therefore has a single positive solution for $\alpha$ for any given value of $\beta$.  While we would need to solve the equation numerically for general values of $\beta$, we can derive approximate solutions analytically for very large and very small values of $\beta$.

Before we do this, however, let us first rewrite this equation in an equivalent form.  We start by changing variables in the first term to $x=\lambda-\frac{1}{2}$, obtaining
\begin{eqnarray*}
&&2\sum_{k=1}^{\infty}(-1)^{k}\realpart\int_{1}^{\infty}dx\,\frac{e^{2\pi ikx}}{\sqrt{x^2+\alpha^2-\tfrac{1}{4}}}+\frac{1}{\alpha} \cr
&&-\ln\left (1+\sqrt{\alpha^2+\tfrac{3}{4}}\right )=\ln{\beta}.
\end{eqnarray*}
We now note that, as a function of $x$, the integral is analytic in the entire complex plane, except for a branch cut.  For $\alpha<\frac{1}{2}$, this branch cut can be chosen to be on the real axis and in the interval $-\sqrt{\frac{1}{4}-\alpha^2}<x<\sqrt{\frac{1}{4}-\alpha^2}$.  For $\alpha>\frac{1}{2}$, on the other hand, the branch cut may be chosen to lie along the imaginary axis.  In either case, we may integrate this function over a large quarter circle centered at the point, $x=1$, in the complex plane and with one of the radii along the positive real axis and the other parallel to the positive imaginary axis and obtain zero since we will always avoid the branch cut.  The contribution from the circular arc will vanish as we increase the radius to infinity since the integrand decreases exponentially as we do so.  This leaves only contributions from the radii.  The contribution from the radius along the real axis is just the integral that appears in the equation.  This means that the contribution from the radius parallel to the imaginary axis is equal to this integral.  We may therefore write
\begin{eqnarray*}
&&\realpart\int_{1}^{\infty}dx\,\frac{e^{2\pi ikx}}{\sqrt{x^2+\alpha^2-\tfrac{1}{4}}} \cr
&&=\realpart\left [i\int_{0}^{\infty}dx'\,\frac{e^{-2\pi kx'}}{\sqrt{(1+ix')^2+\alpha^2-\tfrac{1}{4}}}\right ].
\end{eqnarray*}
Note that we dropped a factor of $e^{-2\pi ik}$; since $k$ is an integer, this factor is always equal to $1$.  If we substitute this back into the equation, we find that the sum on $k$ is just a geometric series with a common ratio of $-e^{-2\pi x'}$.  We may therefore perform the summation, obtaining
\begin{equation}
I(\alpha)+\frac{1}{|\alpha|}-\ln\left (1+\sqrt{\alpha^2+\tfrac{3}{4}}\right )=\ln{\beta},
\end{equation}
where
\begin{eqnarray}
I(\alpha)=2\realpart\left [\int_{0}^{\infty}dx'\,\frac{-i}{\sqrt{(1+ix')^2+\alpha^2-\tfrac{1}{4}}}\frac{1}{e^{2\pi x'}+1}\right ]. \nonumber \\ \label{Eq:I_Integral}
\end{eqnarray}
We can see that the integral $I(\alpha)$ converges for all values of $\alpha$; the integrand is analytic everywhere on the interval of integration and decreases exponentially for large $x'$.  At large values of $\alpha$, we can show that this integral falls off as $\alpha^{-3}$.  We first note that the integral is dominated by small values of $x'$ due to the Fermi occupation factor-like expression.  With this in mind, we may pull out a factor of $\alpha$ from the square root, obtaining
\begin{equation*}
-2\realpart\left [i\int_{0}^{\infty}dx'\,\frac{1}{\alpha}\left [1+\frac{(1+ix')^2-\tfrac{1}{4}}{\alpha^2}\right ]^{-1/2}\frac{1}{e^{2\pi x'}+1}\right ].
\end{equation*}
Since $\alpha$ is large, we now have a small parameter with respect to which we may perform an expansion of the square root.  The constant term in this expansion gives no contribution, since the total result will be purely imaginary.  The lowest-order non-zero contribution will, in fact, be given by
\begin{equation*}
-\frac{2}{\alpha^3}\int_{0}^{\infty}dx'\,\frac{x'}{e^{2\pi x'}+1}=-\frac{1}{24\alpha^3}.
\end{equation*}
We see that this term is of the order $\alpha^{-3}$, as asserted earlier.

We may derive a good closed-form approximation to $I(\alpha)$ as follows.  Let us first expand the square root in the integrand in powers of $x'$.  To the lowest non-vanishing order, we obtain
\begin{equation*}
I(\alpha)\approx -2\int_{0}^{\infty}dx'\,\frac{x'}{\left (\tfrac{3}{4}+\alpha^2\right )^{3/2}}\frac{1}{e^{2\pi x'}+1}=-\frac{1}{3(3+4\alpha^2)^{3/2}}.
\end{equation*}
We now rewrite this expression so that its value at $\alpha=0$ matches the exact value of $I(0)$.  Doing so, we obtain
\begin{equation}
I(\alpha)\approx -\frac{2}{[(-\tfrac{1}{2}I(0))^{-2/3}+4\cdot 6^{2/3}\alpha^2]^{3/2}}.
\end{equation}
If we were to plot this expression alongside the exact expression for $I(\alpha)$, then we would see that it follows the exact expression very closely.  In fact, if we use this expression to solve Equation \eqref{AF_SC_Equation_Alt}, then the solution that we obtain is very close to the solution obtained from the exact $I(\alpha)$.

\section{Derivation of the excitation spectrum} \label{App:VarMF_Ex}
We now present the details of our derivation of the excitation spectrum, given by Eq.\ \eqref{Eq:EXSpectrum}.  As stated before, we begin by constructing a particle-hole excitation of our trial ground state, $b^{\dag}_{\beta}a^{\dag}_{\alpha}\left |0\right >$.  We then find the difference in the expectation value of the Hamiltonian, given by Eq.\ \eqref{Eq:H_FiniteB}, between the excited state and the ground state; this is taken to be the excitation energy.  Throughout this calculation, we assume that the AF order parameter $\Delta>0$.  Let us begin with the quartic terms.  If we take the difference in expectation value of these terms between the excited state and the ground state, we find, after straightforward but tedious application of anticommutation relations and dropping terms that will vanish in the thermodynamic limit, that the contribution to the excitation energy is
\begin{eqnarray}
\delta E_4=\sum_{S}g_S[\mbox{Tr}(S\Sigma_{\alpha\beta})\mbox{Tr}(S\Sigma_{-})+\tfrac{1}{2}\mbox{Tr}(S\Sigma_{+-}S\Sigma_{\alpha\beta})], \nonumber \\ \label{Eq:EXEnergy4}
\end{eqnarray}
where
\begin{eqnarray}
\Sigma_{+-}&=&\sum_{n}[\psi_n^{+}(\br)(\psi_n^{+})^{\dag}(\br)-\psi_n^{-}(\br)(\psi_n^{-})^{\dag}(\br)], \\
\Sigma_{\alpha\beta}&=&\int d^2\br\,[\psi_\alpha^{+}(\br)(\psi_\alpha^{+})^{\dag}(\br)-\psi_\beta^{-}(\br)(\psi_\beta^{-})^{\dag}(\br)]. \nonumber \\
\end{eqnarray}
We may now evaluate the sums and integrals in the above expressions, obtaining
\begin{equation}
\Sigma_{+-}=\frac{m^*}{2\pi}\omega_c(Y1_2\sigma_z s_z+\tau_z 1_2 s_z)
\end{equation}
and
\begin{widetext}
\begin{eqnarray}
\Sigma_{\alpha\beta}&=&\tfrac{1}{16}\left (1+s_1\frac{\Delta}{E_{n_1}}\right )(1+\tau_1\tau_z)(1+\sigma_z)(1+s_1 s_z)\theta_{n_1-1+\tau_1}+\tfrac{1}{16}\left (1-s_1\frac{\Delta}{E_{n_1}}\right )(1+\tau_1\tau_z)(1-\sigma_z)(1+s_1 s_z)\theta_{n_1-1-\tau_1} \cr
&-&\tfrac{1}{16}\left (1-s_2\frac{\Delta}{E_{n_2}}\right )(1+\tau_2\tau_z)(1+\sigma_z)(1+s_2 s_z)\theta_{n_2-1+\tau_2}-\tfrac{1}{16}\left (1+s_2\frac{\Delta}{E_{n_2}}\right )(1+\tau_2\tau_z)(1-\sigma_z)(1+s_2 s_z)\theta_{n_2-1-\tau_2}. \nonumber \\
\end{eqnarray}
\end{widetext}
Here, $\theta_n$ is $0$ if $n<0$, or $1$ otherwise.  We may now evaluate the traces in Eq.\ \eqref{Eq:EXEnergy4}.  Doing so, and using the fact that
\begin{equation}
\frac{m^*}{4\pi}\omega_c(g_{A_{1g}}+g_{A_{2u}}+4g_{E_K})Y=\Delta,
\end{equation}
we find that
\begin{eqnarray}
\delta E_4&=&\Delta^2\left (\frac{1}{E_{n_1}}+\frac{1}{E_{n_2}}\right ) \cr
&+&\frac{m^*}{4\pi}\omega_c(g_{A_{1g}}+g_{A_{2u}}-4g_{E_K})(\tau_1 s_1-\tau_2 s_2). \nonumber \\
\end{eqnarray}
We now consider the quadratic terms.  One of these terms simply gives us the single-particle ``auxiliary spectrum'', while the other is the quadratic term that we chose to consider as part of the interaction term.  We find, upon application of anticommutation relations as before, that the contribution from these terms to the excitation energy is
\begin{equation}
\delta E_2=E_{n_1}+E_{n_2}-\Delta\mbox{Tr}(1\sigma_3 s_z\Sigma_{\alpha\beta}).
\end{equation}
Upon evaluating the trace, we obtain
\begin{equation}
\delta E_2=E_{n_1}+E_{n_2}-\Delta^2\left (\frac{1}{E_{n_1}}+\frac{1}{E_{n_2}}\right ).
\end{equation}
Combining these two contributions, we arrive at Eq.\ \eqref{Eq:EXSpectrum}.

\end{document}